\documentclass[preprint2]{aastex61}

\usepackage{epsf}
\usepackage{graphicx}
\usepackage{lineno}
\usepackage{threeparttablex} 
\usepackage{amsmath}    

\newcommand{\vv}{\mbox{\em V }}
\newcommand{\ii}{\mbox{\em I }}
\newcommand{\bb}{\mbox{\em B }}
\newcommand{\rr}{\mbox{\em R }}
\newcommand{\uu}{\mbox{\em U }}

\newcommand{\bmv}{\mbox{\em B-V }}
\newcommand{\bmi}{\mbox{\em B-I }}
\newcommand{\vmi}{\mbox{\em V-I }}

\newcommand{\umb}{\mbox{\em U-B }}
\newcommand{\umv}{\mbox{\em U-V }}

\newcommand{\feh}{\mbox{\rm [Fe/H]}}
\newcommand{\ebv}{\mbox{\em E(B-V) }}
\newcommand{\evi}{\mbox{\em E(V-I) }}

\providecommand{\e}[1]{\ensuremath{\times 10^{#1}}}

\begin{document}

\title {A Photometric Study of the Outer Halo Globular Cluster NGC$\,$5824}

\author{A.R. Walker}
\affiliation{Cerro Tololo Inter-American Observatory, National Optical Astronomy Observatory, Casilla 603, La Serena, Chile}

\author{G. Andreuzzi}
\affiliation{Fundaci\'{o}n Galileo Galilei - INAF, Bre\~{n}a Baja, La Palma, Spain}
 
\author{C.E. Mart\'{i}nez-V\'{a}zquez}
\affiliation{Instituto de Astrof\'{i}sica de Canarias, Calle Via Lactea, E-38205, La Laguna, Tenerife, Spain}
 \affiliation{Universidad de La Laguna, Dpto. Astrof\'{i}sica, E-38206, La Laguna, Tenerife, Spain}

\author{A.M. Kunder}
\affiliation{Leibniz-Institute f\"{u}r Astrophysik Potsdam (AIP), Ander Sternwarte 16, D-14482, Potsdam, Germany}
 \affiliation{Saint Martin's University, Old Main, 5000 Abbey Way SE, Lacey, WA 98503, USA}
 
\author{P.B. Stetson}
\affiliation{Dominion Astrophysical Observatory, Herzberg Institute of Astrophysics, National Research Council, Victoria, British Columbia V9E 2E7, Canada}

\author{S. Cassisi}
\affiliation{INAF-Osservatorio Astronomica di Collurania, via M. Maggini, 64100 Teramo, Italy}

\author{M. Monelli}
\affiliation{Instituto de Astrof\'{i}sica de Canarias, Calle Via Lactea, E-38205, La Laguna, Tenerife, Spain}
 \affiliation{Universidad de La Laguna, Dpto. Astrof\'{i}sica, E-38206, La Laguna, Tenerife, Spain}

\author{G. Bono}
\affiliation{Departimento di Fisica, Universit\'{a} di Roma Tor Vergata, via della Ricerca Scientifica 1, 00133, Rome, Italy}

\author{M Dall'Ora}
\affiliation{INAF-Osservatorio Astronomica di Capodimonte, salita Moiariello 16, 80131, Napoli, Italy}

\author{A.K. Vivas}
\affiliation{Cerro Tololo Inter-American Observatory, National Optical Astronomy Observatory, Casilla 603, La Serena, Chile}

\correspondingauthor{A.R. Walker}
\email{awalker@ctio.noao.edu}



\begin{abstract}
Multi-wavelength CCD photometry over 21 years has been used to produce deep color-magnitude diagrams together with
light curves for the variables in the Galactic globular cluster NGC 5824.  Twenty-one new cluster RR Lyrae stars are identified, bringing the total to 47, of which 42 have reliable periods determined for the first time.    The color-magnitude diagram  is matched using BaSTI isochrones with age of $13$~Gyr. and
reddening is found to be $E(B-V) = 0.15 \pm0.02$; using the period-Wesenheit relation in two colors the distance modulus is 
$(m-M)_0=17.45 \pm 0.07$ corresponding to a distance of 30.9 Kpc.    
The observations show no signs of populations that are significantly younger than the $13$~Gyr stars.  The width of the red giant branch does not allow for a spread in $\feh$ greater than $\sigma = 0.05$ dex, and there is no photometric evidence for widened or parallel sequences.
The $V, c_{UBI}$ pseudo-color magnitude diagram shows a bifurcation of the red giant branch
that by analogy with other clusters is interpreted as being due to differing spectral signatures of the first (75\%) and second (25\%) generations of stars whose age
difference is close enough that main sequence turnoffs in the color-magnitude diagram are unresolved.    The cluster main sequence is visible against the background out to a radial distance of $\sim17$ arcmin.  We conclude that NGC 5824 appears to be a classical Oosterhoff Type II globular cluster, without overt signs of being a remnant  of a now-disrupted dwarf galaxy.

\end{abstract}


\section{Introduction}

NGC 5824 (C1500-328) is an outer halo globular cluster (GC) that until recently has received little attention, at least in part due to the combination of its considerable distance from the Sun and a high central concentration.   Together these make ground-based observations of individual stars difficult.   A visual inspection by \citet{i1} refers to the cluster as being very condensed, and the \citet{h1} database lists a central luminosity of $4.1\e{4}L_{\sun}$ per cubic parsec, ranking in the highest 20\% of Galactic GCs.     \citet{h1} shows that after NGC 2419, NGC 5824 is the most massive GC more distant than 20 kpc from the Galactic center, and it is the 14th most luminous GC in the Galaxy.  NGC 5694 and NGC 6229 are two clusters at similar distances from the Galactic center, but are a factor 2-3 less luminous than NGC 5824.  

With Galactic coordinates $l = 332\overset{\degr}{.}56$,  $b = 22\overset{\degr}{.}07$ the foreground star contamination from the Galactic bulge is substantial, however the reddening is moderate, with \ebv = 0.13 \citep{z1} or \ebv = 0.15 \citep{s3}, assuming $R = 3.1$ in each case.  \citet{k1} found a spectral type of F5, thus NGC 5824 is metal-poor.  This result has been confirmed by more recent work, in particular \citet{d1}, who derive a metallicity of
$\feh = -2.01\pm0.13$ from spectroscopy of the calcium triplet (CaT) absorption lines, and \citet{r2} who in a high-resolution spectroscopic study of 26 bright red giant branch (RGB) stars found that $\feh = -1.94 \pm 0.02 $ (statistical)$ \pm 0.10$ (systematic), using Fe II lines.   In addition \citet{r2} find no evidence for variation in $\feh$ at the 0.08 dex level in their sample of stars, noting that the stars for which \citet{d1} found a small dispersion were a fainter group than the 26 bright RGB stars that are common to both investigations.    However \citet{r2} find an internal dispersion of 0.28 dex in [Mg/Fe] and point out that this behavior is observed in several other luminous, metal-poor clusters.  Two stars of their sample were observed with Magellan MIKE and show the expected  correlation of [Mg/Fe] with [O/Fe], and anti-correlation of [Mg/Fe] with [Na/Fe] and [Al/Fe].  So although the light-element observations for a substantial sample of stars in NGC 5824 are still lacking, it may be possible to use the [Mg/Fe] measurements as a proxy to split the first and second generation stars.   \citet{r2} also find that a single star of their sample shows substantial $s$-process enhancement; for the other 25 stars they state that $n$-capture abundance patterns are essentially identical and  consistent to those for field stars  with equivalent stellar parameters and metallicities. 

The first color-magnitude diagram (CMD) for NGC 5824 was photographic \citep{h2} and only just reached to the horizontal branch (HB), however the data were of sufficient quality that 10 RR Lyrae variable stars (RRL) could be measured photometrically, resulting in a distance of 25 kpc from the Sun by adopting $M_{\vv}(HB) = 0.6$.  It was noted from the slope of the RGB that the cluster appeared to be metal-poor with a possible blue horizontal branch (BHB).
This was confirmed in the first comprehensive photometric study of NGC 5824, where \citet{c1} used a small-format CCD to measure 285 stars to  $\vv = 20.5$.   Assuming a reddening \ebv = 0.13, they found an apparent modulus of $18.0 \pm 0.2 $ mag and a color for the RGB at the level of the HB of $(\bmv)_{0,g} = 0.77$ corresponding to \feh = -1.7, and confirmed that the cluster has a well-populated BHB.  A deeper CMD  to $\vv \sim 22$ by \citet{b1} showed a possible extreme BHB component (EHB stars) and, while demonstrating some photometric calibration differences with \citet{c1},  derived rather similar cluster parameters, finding
from $\vv_{HB} = 18.45 \pm 0.05$ and $(\bmv)_{0,g} = 0.73 \pm 0.03$ that \feh = -1.85, adopting the \citet{z1} reddening.
A relatively shallow HST \bb and \vv CMD \citep{P8} shows no non-canonical sequences, and neither did the analysis of archival HST observations by \citet{s2}.

Study of NGC 5824 languished between 1995-2008, although several investigators using NOAO and ESO telescopes observed the cluster photometrically in this period; these images can be found in the relevant archives, and are part of the present analysis.  During the same period there was a paradigm shift for globular clusters, with the discovery of association between some clusters and dwarf galaxies past and present, and with many of the 
more massive clusters in particular demonstrating a variety of non-canonical behavior such as metallicity spreads and/or multiple populations 
(see e.g. \citealt{p1}, \citealt{g1}, \citealt{m1}, \citealt{re15}).  In this context \citet{n1} pointed out the possible association of NGC 5824
with the Cetus Polar Stream, for which \citet{p12} finds an orbit almost perpendicular to the northern-sky Disk of Satellites, part of the Vast Polar Structure (VPOS) of satellite galaxies, globular clusters and streams discussed by these authors.  \citet{lm}
exclude NGC 5824 from any association with the Sagittarius dwarf galaxy that  is presently colliding with the Milky Way.
Notably, \citet{g2} included NGC 5824 in their wide-field photographic study searching for tidal tails around a sample of 12 globular clusters.   They demonstrate the good fit of a power law with exponent $-2.2 \pm 0.1$ over the whole range of radius out to a nominal $r =  45$ arcmin, although the high stellar background  means that error bars are large for the outer data points (those with $r > 13$ arcmin).  Indeed, later work by \citet{cb12} shows evidence for a profile truncation starting at about 10 arcmin, and \citet{cb14} find no associated extra-tidal structure; see however the discussion by \citet{d1} in the context of NGC 5824 as being the remnant of a dwarf galaxy that was accreted by the Milky Way several Gyr ago and is now disrupted.  An independent study of the structural properties of 26 GC using a combination of HST and ground-based imaging \citep{mi13} fits King and Wilson models to NGC 5824, the limiting radii derived are very different, 8.79 and 40 arcmin respectively.

In this paper we present a deep and accurate CMD reaching to $\vv \sim24$ from extensive observations on several telescopes over more than two decades, and which we compare to BaSTI (\citealt{p2,p3}) isochrones.   The data set includes observations in $UBVRI$ filters, albeit with many fewer $U$ and $R$ observations.  A few observations were made in other filters and these have been transformed as described below.
We have increased the number of variable stars to 59, of which 47 appear to be RRL cluster members, with in addition one certain and another probable Type II Cepheid, four SX Phe stars, and five eclipsing binaries.  An additional RRL appears to be a very distant halo field star that happens to be in the field of view of NGC 5824.  We have derived positions and mean photometry for all of them, and periods for all except six stars; although a few of these 53 periods are marked (Appendix 1) as not being reliable.   We subsequently calculated light-curve parameters for the RRL in order to help elucidate the nature of NGC 5824.
   
The arrangement of this paper is as follows:  In Section 2 we describe the observations and the data reduction procedures; Section 3 contains color-magnitude and color-color plots and we derive the cluster reddening and metallicity; in Section 4 the RRL variable stars are presented, pulsational quantities derived  and the distance to the cluster determined; then we follow in Section 5 with a brief description of other variables discovered.  In the final section we present a discussion and conclusions, followed by notes on individual variables in an appendix.

\section{Observational Data and Reduction Procedures}

The observations of NGC 5824 consist of 1818 calibrated CCD images from the database maintained by one of us (PBS), obtained in 26 observing runs over a period of 21 years from 1995 March until 2016 February, as listed in Table 1.  The 2008-2009 CTIO 4m Mosaic II observations were taken in time assigned for programs 2008A-0289 and 2009A-0116 to A. Walker (P.I.), and DECam time was provided similarly under program 2013A-2101.  The methodologies  used to produce photometry from such diverse data were similar to those used in the Homogenous Photometry series as described in detail by \citet{st2} and are based on the DAOPHOT/ALLSTAR/ALLFRAME suite of programs \citep{st87, st94}, which are available from P.B. Stetson. The reductions produced a catalog containing 91,779 stars,\footnote{The photometric catalog from the ALLFRAME reduction, together with the time-series ALLSTAR photometry for 59 variable stars, are available from the Canadian Astronomy Data Center (CADC) at 
http://www.cadc-ccda.hia-iha.nrc-cnrc.gc.ca/en/community/STETSON/homogeneous/.}  that tabulates the mean calibrated photometry, astrometry, and an indication of variability established using an updated version of the Welch-Stetson method \citep{ws} run over the individual reduced exposures.   

\begin{table*}
    \begin{scriptsize}
    \caption{Observations}
    \label{tab:one}
    \begin{tabular}{ccccccc}
    \hline
    \hline    
     
     No. & ID   &   Dates            &           Telescope   &      Camera  & No. Images & Notes \\
\hline
     1 & 	emmi2 &       1995 Mar 07-10      &          ESO NTT 3.6m   &     EMMI      &      19  & \\
     2 & 	bond6  &      1998 Apr 16-22        &        CTIO 0.9m      &     Tek2K3   &       4   & \\
     3 & 	wfi3  &       1999 Mar 14-19        &        ESO/MPI 2.2m   &     WFI       &      80  & \\
     4 & 	wfi10: &      2000 Jul 06-12        &        ESO/MPI 2.2m   &     WFI       &     112  & \\
     5 & 	mzocc &      2000 Jul 25-27        &        ESO/MPI 2.2m   &     WFI       &     104  &  \\
     6 & 	susi2a  &     2001 Jul 20-24        &        ESO NTT 3.6m   &     SUSI2     &      30  & \\
     7 & 	wfi6   &      2002 Feb 18-21        &        ESO/MPI 2.2m   &     WFI       &      88  & \\
     8 & 	susi03may &   2003 May 30-31       &         ESO NTT 3.6m   &     SUSI      &      18  & \\
     9 & 	west1  &      2005 Feb 11-17      &          CTIO 0.9m      &     Tek2K3   &      13  & a \\
    10 & 	B05may  &     2005 May 11-12     &           CTIO 4.0m      &     Mosaic2   &     192  & \\
    11 & 	fors20602 &   2006 Feb 23-28      &          ESO VLT 8.0m   &     FORS2     &       8  &  \\
    12 & 	fors20605 &   2006 May 28-30      &          ESO VLT 8.0m   &     FORS2     &       8  & \\
    13 & 	efosc08a &    2008 Apr 16-28      &          ESO NTT 3.6m   &     EFOSC     &      26  & \\
    14 & 	andi4 &       2008 Apr 17-May 14   &         CTIO 1.3m      &     ANDICAM   &      63  &  \\
    15 & 	may08 &       2008 May 02-03       &         CTIO 4.0m      &     Mosaic2   &     304  &  \\
    16 & 	soar08may &   2008 May 02-05       &         SOAR 4.1m      &     SOI       &      70  &  \\
    17 & 	jul08 &       2008 Jul 27-29       &         CTIO 4.0m      &     Mosaic2   &       8  &  \\
    18 & 	aug08 &       2008 Aug 26-28      &          CTIO 4.0m      &     Mosaic2   &     120  &  \\
    19 & 	wfi46 &       2010 Feb 15-19      &          ESO/MPI 2.2m   &     WFI       &     104  &  \\
    20 & 	Y1008 &       2010 Aug 13-18       &         CTIO 1.0m      &     Y4KCam    &       2  &  \\
    21 & 	fors1103 &    2011 Mar 01-03        &        ESO VLT 8.0m   &     FORS2     &      84  &  \\
    22 & 	lee2 &        2011 May 29-Jun 03    &        CTIO 4.0m      &     Mosaic2   &     168  & a \\    
        23 & 	D13 &         2013 May 05, Jul 12   &        CTIO 4.0m     &    DECam        &    5  & b,c \\  
    24 & 	lcogt1 &      2014 Jan 19-Apr 30    &        Sutherland 1.0m &  CCD          &   88  &  \\
    25 & 	D14  &       2014 Feb 26, Mar 08-10  &      CTIO 4.0m     &    DECam        &   46  & b,c  \\
    26 & 	decam1  &    2016 Feb 24           &        CTIO 4.0m     &    DECam        &   54  & b,c \\
\hline  
\end{tabular}
\begin{tablenotes}
\item Notes. --
(a) Str\"{o}mgren filters
(b) SDSS filters
(c) We processed only the one (in a few cases two) DECam CCDs  containing the cluster
\end{tablenotes}
\end{scriptsize}
\end{table*}

 From the 26 observing runs considered here, our data include 117 datasets
obtained with $UBVRI$ filters, where a dataset is defined as
one night's data from one CCD when the weather was photometric, or one or more
night's data from one CCD when the weather was non-photometric.  These observations were
transformed to the  Johnson-Cousins $UBVRI$ system as defined by the 
standard stars established by \citet{lan92}
by comparison of our instrumental
magnitudes to the standard values provided by Stetson (catalog
revision 2013 April 22).  Stetson's list includes 442 stars with photometry
published by Landolt (1973, 1992) also having additional observations in Stetson's
archive; the listed magnitude values for these stars are the weighted average of
Landolt's and Stetson's.  Stetson's list also includes some 117,360 additional
secondary standards, not observed by Landolt, whose observed magnitudes have
been transformed to the Landolt system as described in the aforementioned
papers.  When we compare our derived calibrated magnitudes from those datasets
to Landolt's published values, we find root-mean-square differences of 0.025 mag
per star based on 166 measurements of 14 different standards in $U$, 0.022 mag
(759 measurements of 73 standards) in $B$, 0.017 mag (1452 measurements of 92
stars) in $V$, and 0.021 mag (488 measurements of 43 stars) in $I$.  No
measurements of Landolt's original standards were obtained in the $R$ filter. 
When our calibrated magnitudes are compared to Stetson's list, we find r.m.s.
differences of 0.010 mag (46,552 measurements of 5,365 standards) in $U$, 0.011
mag (122,123 of 13,348) in $B$, 0.010 mag (191,569 of 15,512) in $V$, 0.003 mag
(34,923 of 3,591) in $R$, and 0.009 mag (124,169 of 9,279) in $I$.  

The source data includes observations made in SDSS $g$, $r$, and $i$ filters and also in Str\"{o}mgren $b$ and $y$ filters  (nos. 9 \& 22 in Table 1).  These were transformed to Landolt $B, R, I$  and $B, V$ respectively.  The SDSS $r$ and $i$ filters are similar to the Landolt \rr and \ii bandpasses and therefore we have treated the SDSS $r$ and $i$ observations as if they were normal \rr and \ii observations.   However the SDSS $g$ filter is almost exactly half-way between $B$ and $V$, we chose to express SDSS $g$ as \bb magnitude plus a function of \bmv color, with linear term $\sim 0.5$ and quadratic term $\sim 0.02$, and then these relations were inverted to predict Landolt \bb from SDSS $g$.  For non-variable stars the prediction is quite robust with star-to-star residuals  $< 0.02$ mag. rms.  However, for variable stars the correction is applied using mean color, not instantaneous color, and as a consequence, the mean \bb magnitude for a variable star predicted from SDSS $g$ should be good, but the amplitude will be underestimated. 
In contrast, the Str\"omgren $b$ filter is somewhat shorter than Johnson
$B$, so transformed $b$-band observations will overestimate $B$-band amplitudes by
a small amount.  Conversely, Str\"{o}mgren $y$ tracks \vv much more closely, with small color terms.    
In  this paper we discuss only the \vv amplitudes, so avoid this issue.

The cluster center is estimated to be at (x,y) = (+9,-2) arcsec on our catalog coordinate system, which corresponds to RA, Dec of $15^h 03^m 59\overset{s}{.}22,  -33^d 04^m 07\overset{s}{.}3$, within $2-3$ arcseconds of the cluster center measured by \citet{s2,h1}.  The center of the cluster is very crowded and we wish to produce a sub-catalog  with excellent photometry and dominated by cluster stars, for calculation of cluster parameters and for comparison with isochrones.   We therefore proceeded to make a number of selection cuts on the main catalog.

     \begin{figure}
  \includegraphics[width=\columnwidth]{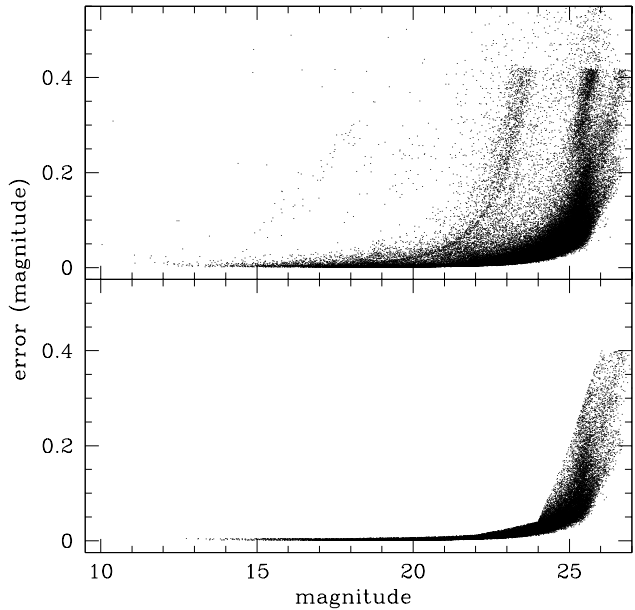}
  \caption{Photometric error as a function of magnitude for the \bb band data.  The upper panel plots all stars with \bb  band  photometry, the lower panel shows the stars remaining after the error cuts listed in Table 2.}
     \label{fig-one}
  \end{figure}
 
 The principal issues to be considered are the photometric depth as a function of radial distance from the cluster center, and the effects of crowding as the cluster center is approached.  
 The spatial coverage is essentially complete out to a radius of 22.7 arcmin if we consider stars with at least one calibrated observation in $V$ and also one of $B$ or $I$.  Coverage is partial out to (our most distant star) 33.9 arcmin.   More scientifically meaningful, if we consider only stars with at least 5 calibrated observations in all of  $B, V, I$, then spatial coverage is complete out to 19.6 arcmin and partial out to 32.3 arcmin.  The great majority of the cluster stars are closer to the cluster center than any of these radii (see 3.1).

 Figure 1 shows the photometric error as a function of magnitude for the \bb passband, and plots for the other bands are similar.  The multiple sequences in the upper panel are due to the variable depth as a function of sky position, clearly demonstrating the non-homogenous nature of the data set.  From plots such as these it is straightforward to apply a cut on error as a function of magnitude, here we reject stars with errors greater than an envelope formed by limits defined at one magnitude intervals, as shown in Table 2 and with the results depicted in the lower panel of Figure 1.  For all the brighter stars, those with photometric error greater than 0.005  mag are discarded, at the level of the cluster sub-giant branch (SGB) the error cut is still only $\sim0.01$ mag.   Following these cuts, the distribution of photometric error as a function of magnitude for each filter is a reasonably smooth progression.  We also made cuts of the DAOPHOT photometric fitting parameters $\it{chi}$ and $\it{sharp}$, to remove non-stellar objects such as galaxies, CCD defects, and unresolved double stars.  We rejected objects with $\it{sharp} $ $< -0.2$ and $> 0.2$, and $\it{chi}$ $ < 0.7$ and $>2$.   Given the often complex background for any given star, the discriminatory power of these parameters is less incisive than in uncrowded fields, where rather tighter limits would normally be applied, therefore these cuts can be considered supplementary to those based on the measured photometric errors.  Considering only stars brighter than magnitude 22, approximately the level of the main sequence turnoff, these cuts removed two-thirds of the \uu band measurements and approximately half of the measurements in each of the other bands.   Because this process has the side-effect of efficiently discarding variable stars, we separately add back those for which we determined periods in the plots we show below, realizing that some variables for which we do not determine periods will remain discarded, since we have no reliable way of separating them from constant stars with poor photometry.   We will discuss the variable stars in detail later, so suffice it to say here that a total of 94 stars were selected as being possibly variable and so subject to further scrutiny.

  \begin{table}
    \begin{scriptsize}
    \caption{Error limits in magnitudes for each filter, as a function of magnitude}
    \label{tab:err}
    \begin{tabular}{cccccc}
    \hline 
    \hline
    Mag. & \uu & \bb & \vv & \rr & \ii \\
    \hline
    10  &  0.005  &  0.005  &  0.005  &  0.005  &  0.005\\
    16  &  0.005 & 0.005 & 0.005 & 0.005 & 0.005\\
    17  &  0.006  &  0.005  & 0.005   &  0.005   & 0.005\\   
    18  &  0.007 &  0.005  &  0.005  &  0.005  &  0.006\\  
    19 &  0.010  &  0.006  &  0.007  &  0.005  &  0.008\\  
    20 &  0.015  &  0.007  &  0.008  &  0.006  &  0.012\\ 
    21  &  0.03  &  0.009  &  0.01  &  0.01  &  0.02\\  
    22  &  0.05  &  0.012  &  0.018  &  0.02  &  0.05\\
     23 &  0.10  &  0.025  &  0.03  &  0.05  &  0.10\\    
     24  &  0.4  &  0.04  &  0.10  &  0.15  &  0.2\\ 
     25   &  0.4  &  0.2  &  0.4  &  0.4  &  0.4\\   
     26  &  0.4  &  0.4  &  0.4  &  0.4  &  0.4\\    
    \hline
    \end{tabular}
    \end{scriptsize}
    \end{table}

\begin{figure}
   \includegraphics[width=\columnwidth]{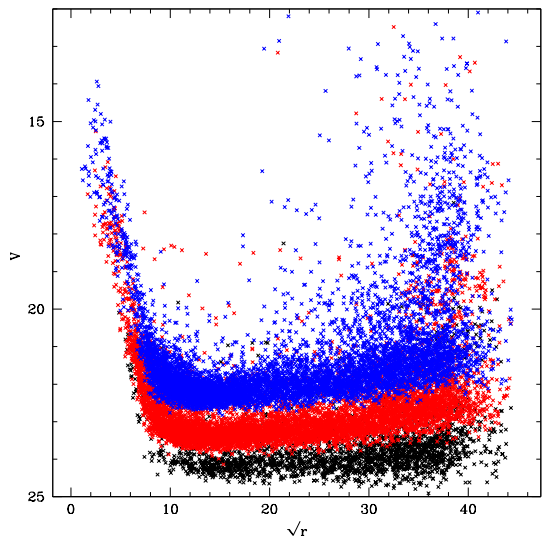}
   \caption{Plot of the $V$ magnitude   against the square root of the radial distance (arcsec) from the cluster center for stars  with three ranges of error in $B-I$ color, using the original (complete) catalog as source.  Color coding is: For stars with $0.09 < \sigma(B-I) < 0.10$ mag   [black]; $0.03 < \sigma(B-I) < 0.04$ mag   [red]; $0.01 < \sigma(B-I) < 0.015$ mag  [blue]}
  \label{fig-two}
\end{figure}

We turn now to considering the effects of crowding, demonstrating in Figure 2 the radial zones of good photometry by plotting $V$ magnitude as a function of radius, for three ranges of the error in the $\bmi$ color.  Similar plots are obtained for other combinations of magnitude and color.  The quality of the photometry slightly improves inward to a radius of $\sim 80$ arcseconds, due to the number of observations increasing. Inward of this radius the magnitude limit for stars of each quality class rapidly brightens, suggesting that crowding errors set in abruptly at this radius.  Note that the RRL, at $V \sim 18.4$, transition from the best to the poorer quality classes at a radius $\sim 36$ arcseconds.    Outside a radius $\sim 900$ arcseconds, the quality of the photometry deteriorates slowly for most stars, and a significant fraction of the area has much shallower detection limits out to the outermost stars at  $\sim 2000$ arcsec; this aspect is also demonstrated in Figure 1 for the \bb band.

\section{The Color-Magnitude Diagram}

\subsection{Radial distribution}  

Figure 3 shows the observed CMD for a range of eight annuli, in total extending from 1.5 to 25.5 arcmin from the cluster center.   The cluster  main sequence (MS) dominates in the inner annuli, is clearly still visible in the 13.5-16.5 arcmin radius panel, is just visible in the 16.5-19.5 arcmin radius panel, and is absent in the two outermost panels.   The same can be said for the BHB stars.  Our data out to and slightly beyond the 19.5 arcmin radius is still relatively spatially homogenous but becomes slightly photometrically shallower as radius increases (see above).  For cluster RRL confirmed to be members from their position in the CMD, we find only four more distant than a radius of 8.5 arcmin: V46 (8.63 arcmin), V26 (9.82 arcmin), V32 (12.91 arcmin), V29 (19.32 arcmin).    
 
\begin{figure*}
   \includegraphics[scale=0.8]{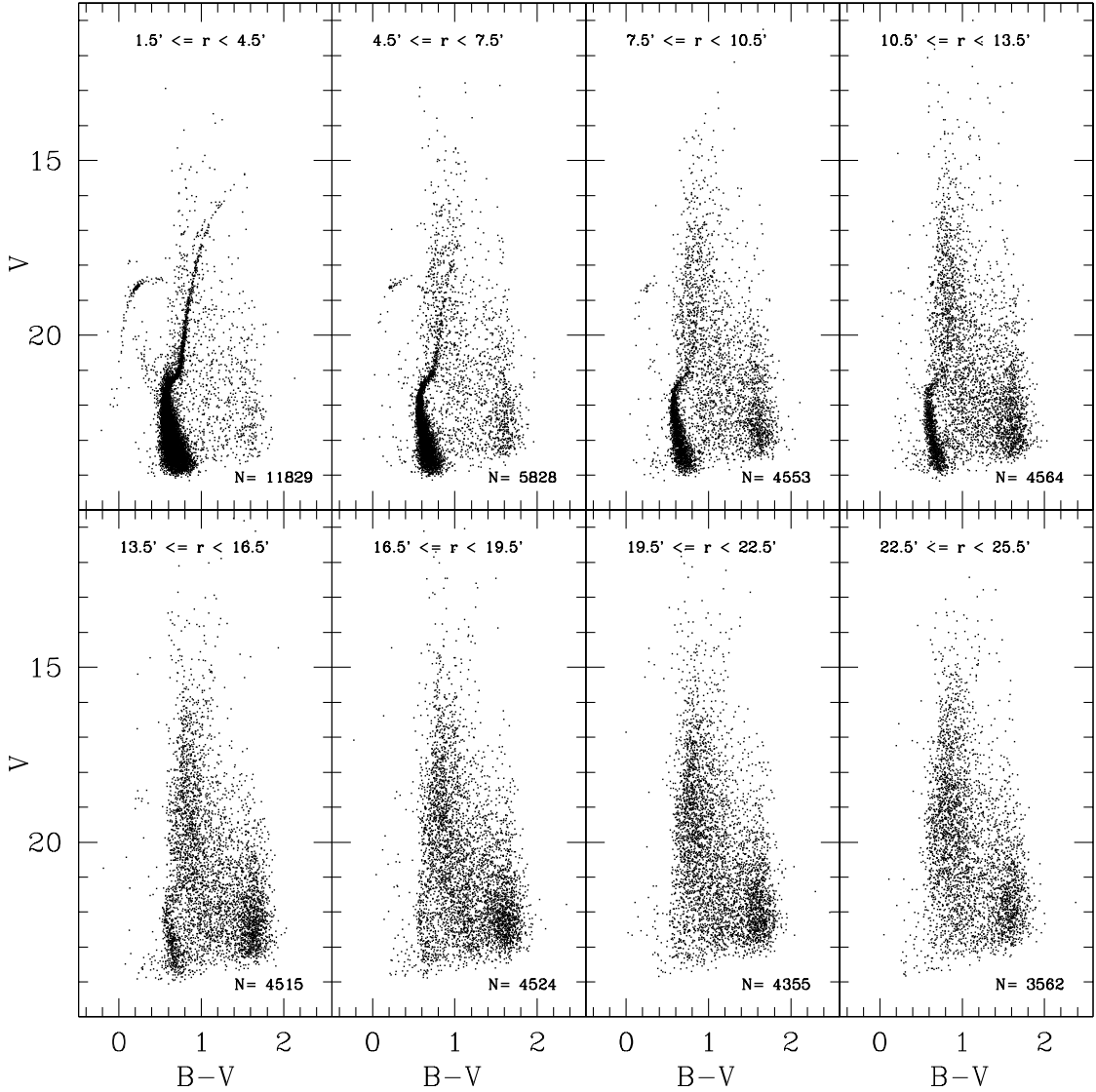}
   \caption{$V, B-V$ ~color-magnitude diagrams for eight annuli centered on NGC 5824,  Each panel lists the inner and outer radii, and the number of stars in the figure.}
  \label{fig-three}
\end{figure*}

As previously mentioned, since the work of \citet{g2} the star density profile of NGC 5824 has been known to be unusual, and as shown by \citet{cb12} a King model is a particularly poor fit to NGC 5824 (and to some other massive GCs in their sample), whereas a power law fit to their data is much better, with a truncation indicated near a radius of 15 arcmin.  Our data, which show the cluster MS persisting into the  $16.5-19.5$ arcmin radius bin would argue for a slightly greater cluster extent, but due to the non-homogenous nature of the present data set at larger radii it is not possible to significantly add to this discussion;  analysis of an homogenous, deep, wide-field data set (e.g., with DECam) is out of scope of this work but needed to make further progress.  \citet{d1} promise such an analysis. 

\begin{figure}
   \includegraphics[width=\columnwidth]{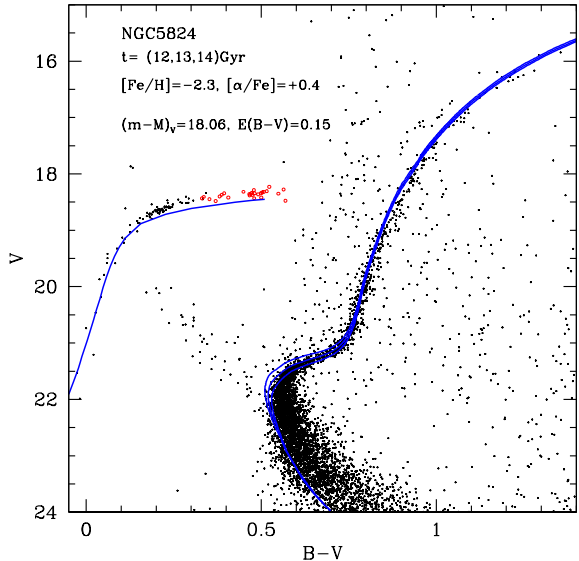}
   \caption{$V, B-V$  CMD for stars in an annulus between 90 and 270 arcsec from the center of NGC 5824.  The data has stars with large errors removed, as described in the text.  RRL stars are plotted as red circles.  BASTI isochrones are overlaid, as  specified on the figure and described in the text. }
  \label{fig-four}
 \end{figure}

\begin{figure}
   \includegraphics[width=\columnwidth]{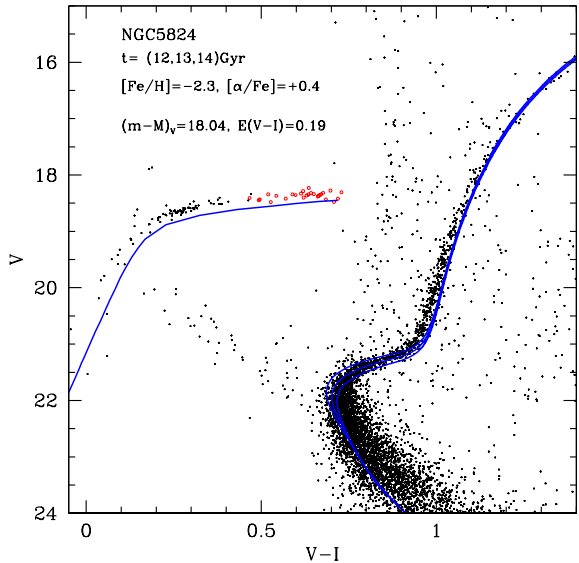}
   \caption{$V, V-I$ CMD, otherwise as for Figure 4.}
  \label{fig-five}
 \end{figure}

\subsection{Comparison with isochrones}

Based on the above discussion and on the experiments described in  Section 2, we decided to use stars lying in an annulus centered on the cluster and with radii $90 - 270 $ arcsec to define the cluster sequences, as the best compromise between the effects of crowding of cluster  stars that strongly impacts photometry for the fainter stars for radii $<90$ arcsec, and the contrast of the cluster sequences against the swarm of field stars and faint unresolved galaxies at a density of $\sim 200$ arcmin$^{-2}$ to $V=24$.    In addition we use the clean catalog with error cuts as shown in Table 2, and further excluded stars with errant values of DAOPHOT $\it{chi}$ and $\it{sharp}$ as previously described.  This results in  a catalog with 16,194 stars, starting from a full-field clean catalog of 83,878 stars.    

The CMD of NGC 5824 has been compared with the BaSTI theoretical framework \citep{p2,p4}
 in order to retrieve an estimate of the cluster age as well as an
independent determination of the cluster distance and reddening. For this comparison, we adopted
the $\alpha-$enhanced version of the BaSTI isochrones (see \citealt{p4} for details).
In order to perform a more detailed comparison between the empirical dataset and the theoretical
framework, the metallicity grid points in the original BaSTI library have been supplemented
with additional stellar models computations for Z=0.0002, Y=0.245 and the same heavy element
distribution already used for the $\alpha-$enhanced BaSTI models. These additional 
stellar models/isochrones are fully consistent with the original BaSTI models as far as concerns
the adopted physical inputs and numerical assumptions.
Comparisons between the best matching theoretical isochrones and the observed CMDs in the $(V, B-V)$
and $(V, V-I)$ photometric planes are shown in Figs. 4 and 5.
The best result is obtained for an apparent distance modulus equal to ${\rm (m-M)_V=18.04}$~mag
and a reddening \ebv = 0.15, while the age of the isochrone that matches the
MS turn-off and the SGB of the observed CMD is equal to $13$~Gyr.  The (internal) error in the age is $\pm 1$ Gyr. 

We note that the [Fe/H] value adopted for the best matching isochrones is about 0.3~dex more metal-poor than spectroscopic measurements. Indeed,
a reasonable - although slightly worse - result can be obtained by using an iron abundance about 0.15~dex larger than the value adopted
for the isochrones shown in Figs. 4 and 5. We think that this small mismatch between the spectroscopic measurements and the estimate based
on the isochrone matching is due to a combination of uncertainty of the spectroscopic calibrations and to residual shortcomings in the color-${\rm T_{eff}}$ relations used for transferring the stellar models from the theoretical plane to the observational ones.

\begin{figure}
   \includegraphics[width=\columnwidth]{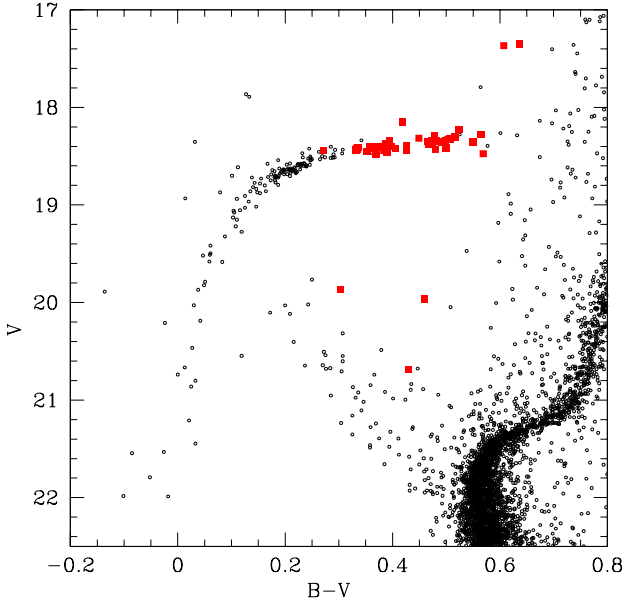}
   \caption{$V, B-V$ CMD for stars in an annulus between 36 and 270 arcsec from the center of NGC 5824.   The plot is restricted to the region of the CCD in the vicinity of the horizontal branch and the blue straggler stars, with all the variable stars with measured periods plotted as filled squares,   Stars with large errors have been removed, as described in the text.}
  \label{fig-six}
 \end{figure}

The distance modulus in the comparison between the observed CMD and the theoretical models has been mainly fixed by the requirement to match
the fainter boundary of the HB stellar distribution. The BaSTI models used in present work have been computed by using the conductive opacity by \citet{po99}, the use\footnote{The computation of a new release of the BaSTI
theoretical framework based on the most recent physical inputs is in progress.} 
of the most updated conductive opacities provided by  \citet{ca07} - when keeping fixed all the other physical inputs - causes a decrease of the ZAHB luminosity by about $\Delta{M_V}\approx0.05-0.07$~mag. It is worth noting that this shift would put the distance modulus obtained from the isochrone fitting procedure in agreement - within $1\sigma$ - with that obtained by using the period-Wesenheit relation (section 4.2).

\subsection{The Horizontal Branch and Blue Straggler Stars}

We show (Figure 6) a CMD of the horizontal branch (HB) and blue straggler (BS) region of the CMD, made from the same cleaned catalog described above, but in this case extending to within 36 arcsec from the cluster center, and out to the same 270 arcsec radius, with the exception that all the identified variable stars with measured periods are plotted.
Figures 4, 5 and 6 all show a tight cluster turn-off region and the evolved stars sequences (SGB, RGB, AGB, HB) are all well-defined.  
 Although the RHB area of the CMD is contaminated with field stars, there are likely a few RHB stars judging by their appearance close to the cluster center (e.g. at radii less than 60 arcsec) where most stars in the vicinity of the cluster CMD sequences statistically belong to the cluster rather than to the field.

Over a wider area, we confirm the description by \citet{s2} who show from HST observations in their Figure 3 that the HB extends to $m_{556} =  22$ (EHB stars), thus the HB extends beyond the so-called "Momany gap" \citep{mo04} that occurs at $\sim$22,000K, as commonly found in the most massive GCs.   We also find, as seen in earlier studies, a prominent clump of BHB stars slightly hotter than the position of the RRL instability strip blue edge.   From simple star counts using the full sample of stars with no error cuts to avoid bias when comparing to the number of RRL, and for stars lying within the annulus with radii 36 and 270 arcsec, there are 32 RRL, $199 \pm 7$ BHB stars and $7 \pm 2$ RHB stars, therefore the HB descriptor introduced by  \citet{le91} is 
 $(B-R)/(B+V+R) = 0.81 \pm 0.04$, where $B,V,R$ are respectively the numbers of BHB, RRL, RHB stars.    Dividing the EHB stars arbitrarily at $\bmv = 0.10$, there are 40 stars bluer and 159 stars redder than this divisor.  The exact evolutionary status of the hottest HB stars in GC  is somewhat controversial (see e.g. \citealt{mo2007}), with spectroscopy essential for interpretation.
 
 We draw attention to a few stars that are bluer and/or brighter than the BHB.  These are non-variables with excellent photometry and, while they may be field stars, an alternative interpretation is that they are BHB stars at the end of their helium-core burning lifetime evolving across the HR diagram to the AGB.  In support of that interpretation we also see that \cite{s2} show several supra-HB stars on their inner (HST) CMD where the field star contamination is low, although it is possible that some of these stars may be RRL scattered into strange positions on the CMD due to their mean magnitudes poorly sampling the light curves.  That explanation cannot be the case for the present many-epoch data set.  There is also one very blue star, 69496, with \vv=19.888, \bmv = -0.136, \umb = -1.001 (see Figure 6), at position RA $15^h 04^m 09 \overset{s}{.}86$, Dec $-33^d 02^m 07\overset{s}{.}3$ (J2000).  These colors correspond closely to an unreddened DA white dwarf with $T_{eff} = 25000K$ \citep{ch93}, and this interpretation seems the most likely.  However the star is only 3.33 arcmin from the cluster center, and the alternative explanation that it is an extremely hot cluster star evolving from the EHB requires spectroscopy for confirmation.  

BS stars are prominent, as previously noted and discussed in detail by \citet{s2}, with approximately 50 in the same annulus as for the HB numbers above, the uncertainty arising from the field contamination of the redder stars.  We discuss below  four SX Phe stars and an RRL star occurring among the BS stars.

\subsection{Metallicity}

We adopt the mean of the \citet{d1} and \citet{r2} metallicity results, $\feh = -1.97 \pm 0.10$, noting that this is identical to earlier measurements to within the errors \citep{h1,ca09}.  Following techniques first introduced by \citet{sk83} and developed by \citet{f91}, we linearized the RGB in the $V, B-V$ CMD between $16<V<21$ by fitting a cubic polynomial, then measuring the color difference between stars and this fiducial, for stars closer to it than 0.05 mag in $B-V$ color.  The distribution is approximately Gaussian with $\sigma = 0.016$ mag.   If all the source of this distribution is assigned to a variation in $\feh$ then it corresponds to $\sigma\feh = 0.10$  \citep{f91}.  This is clearly an upper limit to any metallicity spread and, we can obtain a more realistic estimate  by subtracting in quadrature the photometric errors per star - remembering that we are only considering stars selected to have low photometric errors (see Table 2) - to find that the mean contribution from this source is $\sigma = 0.004$ mag.    
We can also expect small star to star reddening differences and general experience suggests
that small-scale variations are typically at least several percent of the overall
average reddening \citep{be17}, thus a variation of $ 5$ percent of the reddening value would introduce $\sigma \sim 0.007$ mag.   Subtracting these two sources of dispersion in quadrature leaves  $\sigma = 0.006$ corresponding to $\sigma\feh = 0.05.$

\subsection{Reddening}

We took the catalog for the $90 - 270$ arcsec annulus, and carefully examined the individual $\vv, \vmi$ CMD for each quadrant, to check for differential reddening.  We compared the four near-vertical MS for $ 21.5 < \vv < 22.0$, for which the MS width is $\sim 0.08$ mag, and derived a  mean color for the stars within  0.05 mag. of the MS ridgeline.   The mean colors for the stars  in the four photometric boxes were found to be:   $0.738 \pm 0.002$ (NE, 287 stars), $0.736 \pm 0.002$ (NW, 303 stars), $0.738 \pm 0.002$ (SW, 292 stars), $0.740 \pm 0.002$ (SE, 239 stars), and similar results showing no measurable differential reddening above $\sim 0.002$ in \evi were also found when considering smaller areas within the $90 - 270$ arcsec annulus.  The NASA/IPAC dust extinction tool (http://irsa.ipac.edu/applications/DUST/ further confirms that up to a radius of $\sim 10$ arcmin from the center of NGC 5824 the reddening  appears constant, with \citet{s3} finding $\ebv = 0.142 \pm 0.004.$  We further checked by looking at the color of the lower RGB at $\vv = 20.0$, finding the same color in all quadrants, to conclude that any differential reddening on arcmin scales must be $\Delta \ebv \ll 0.01$ mag and so using a mean reddening value for stars in the $90-270$ arcsec annulus is appropriate.  

The actual value for the reddening can be estimated from the boundaries of the instability strip, in particularly the boundary between the BHB and the RRc stars which appears, at least in \bmv color, to be independent of metallicity, at $(B-V)_0 = 0.18$ (\citealt{wal92} \& references therein).  From the Figure 5 data,  there is  a small amount of color overlap between the reddest BHB stars and the bluest RRL, and we estimate the boundary to be $ B-V = 0.33 \pm 0.02$, therefore $E(B-V) = 0.15 \pm 0.02$.

The reddening can also determined from an empirical method originally developed by \citet{st66} that uses  $(B-V)$ colors for RRab variables at minimum light together with terms in period and metallicity, expressed by \citet{wa90} as
\\
$E(B-V) = (B-V)_{min} - 0.24 P - 0.056 \feh - 0.336$  (1)
\\
where the minimum light corresponds to phases between 0.5 and 0.8, period (P) is in days, and the metallicity is on the \citet{zw84} scale.  For redder passbands simple expressions with only the color at minimum light are adequate, with \citet{gu05} and \citet{ku13b} finding
\\
$E(V-R) = (V-R)_{min} - 0.22$    (2)
\\
and
\\
$E(V-I) = (V-I)_{min} - 0.58.$     (3)
\\
With $\feh = -1.97$, we then select only those RRab stars with excellent light curves (V1,V9,V10,V12,V14,V23,V32,V34,V44) and measure the colors at minimum light from the fitted templates.  Following \citet{ku13b} in choice of reddening law when converting from $E(V-I)$ and $E(V-R)$ to $E(B-V)$, the three colors $B-V$, $V-R$ and $V-I$  give a weighted mean of $E(B-V) = 0.15 \pm 0.03$, without including any estimate for systematic errors in the method.  

\begin{table*}
    \begin{scriptsize}
    \caption{Minimun Light Colors for Selected RRab Stars}
    \label{tab:min}
    \begin{tabular}{ccccccccc}
    \hline 
    \hline
    Variable & Period & $B$min & $V$min & $R$min & $I$min & $E(B-V)$ & $E(V-R)$ & $E(V-I)$ \\
    \hline
     V1   & 0.59725  & 19.372   &   18.768   &  18.451  & 18.036  & 0.231  &  0.046  &  0.151  \\
     V9   &  0.67762 & 19.379 & 18.777 &   18.327  & 17.926 &  0.210  &   0.180  &   0.272   \\
    V10   & 0.69472  & 19.271  & 18.750 & 18.343 &  17.888  & 0.124 &  0.137  &   0.283   \\
    V12    & 0.58008 &  19.392  & 18.872  & 18.467  &  18.076 &   0.151  &   0.134  &   0.215  \\
    V14  &  0.71077  & 19.181  & 18.644  &  18.307 &  17.901  &  0.136   &  0.067  & 0.163   \\
    V23  &  0.62808 &  19.373  & 18.872 &  18.359   &  18.031  &  0.120 &  0.244 &  0.261   \\
    V32  & 0.61479  & 19.314  &  18.738  & 18.375  &  18.028  &  0.199  &  0.093   &  0.129   \\
    V34  &  0.79620  & 19.156   & 18.577  &  18.202  & 17.724  & 0.158  &  0.105  &  0.273   \\
    V44  &  0.66460  & 19.319  &   18.651  & 18.312 &  17.875  & 0.279 &    0.070  & 0.197  \\
    \hline
    \end{tabular}
    \end{scriptsize}
    \end{table*}

Other photometric methods available here use the position of the RGB and have strong dependence on the cluster metal abundance, so are not as preferable to use as those discussed above, which in the mean give $E(B-V) = 0.15$, which we adopt.   We can check the consistency of our adopted reddening and metallicity values  by comparing the position and shape of the RGB relative to the HB, and with the 

$(B-V)_{0,g}$, $\Delta V_{1.2}$, $\Delta V_{1.1}$ relations as given by \citet{sl97} then with $E(B-V) = 0.15$ find $\feh = -1.84, -1.81, -1.91$ 
respectively,  not far from our adopted value of $\feh = -1.97$.
 
\subsection{Multi-populations?}

We take advantage of having good photometry as deep as $U \sim 23$  to plot in Figure 7 a $V, U-V$ CMD for a cleaned sample (as described above) of stars in the 90 to 270 arcsec annulus centered on the cluster.  Such diagrams that include the ultraviolet are useful for characterizing multi-population behavior \citep{mil16}.  Specifically, we note that the SGB is narrow and cleanly defined, showing only a single sequence of stars, however the lower RGB is very broad.  All these stars are brighter than $U =21$ for which (Table 2) the error cut is 0.03 mag and the median error is half that.  By $U = 20$ the errors are a factor 2 smaller, thus photometric errors only make a minor contribution to the RGB color spread.  Similarly, any contribution from differential reddening (see discussion above) will contribute less than 0.02 mag to the color spread.   We can proceed further by calculating the $c_{UBI} = (\umb)-(\bmi)$ pseudo-color, which was introduced by \citet{mo13} as a photometric means for identifying multiple stellar populations on the RGB of GCs.   We further pre-clean the stellar sample by use of the $\bmi$ vs. $\umv$ color-color diagram (Figure 8) in which the cluster RGB is clearly differentiated from the more metal-rich foreground stars \citep{bo10,mo13}.  The RGB was fitted by a polynomial of form 
\\
$U-V = 0.2629(B-I)^2 +0.2406(B-I) -0.5014$    (4)
\\
and stars within $\pm0.1$ mag in \umv color of the lfiducial were assigned to the RGB, for $ 0.5 < U-V < 1.5$.

\begin{figure}
   \includegraphics[width=\columnwidth]{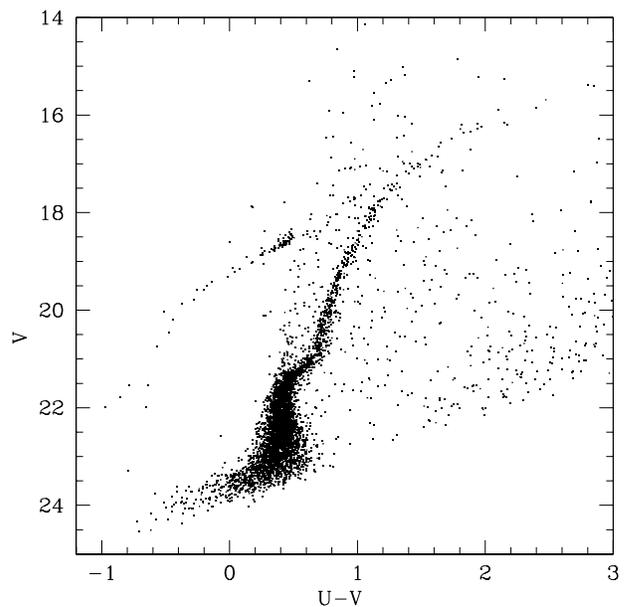}
   \caption{\vv, \umv CMD  for stars within the inner annulus ($90 - 270$ arcsec) with photometric errors less than the limits given in Table 2.}.
     \label{fig-seven}
\end{figure}

\begin{figure}
   \includegraphics[width=\columnwidth]{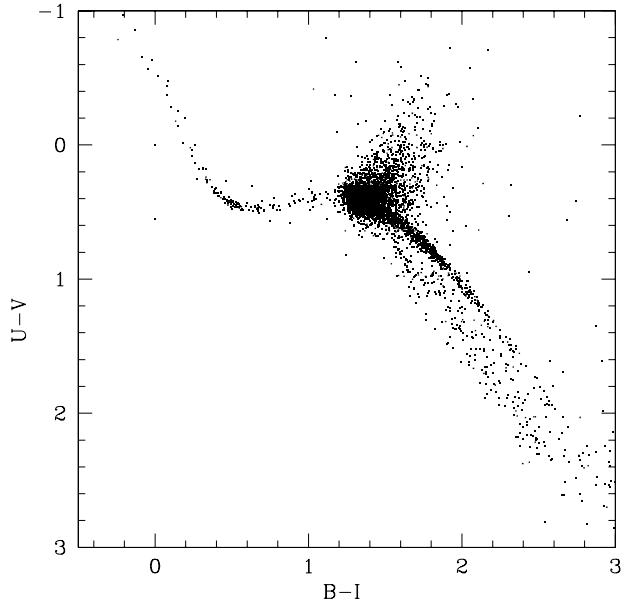}
   \caption{\bmi, \umv color-color diagram for stars within the inner annulus ($90-270$ arcsec). The cluster RGB on the right side of the diagram can be clearly differentiated from field stars.    }
  \label{fig-eight}
\end{figure}

\begin{figure}
   \includegraphics[width=\columnwidth]{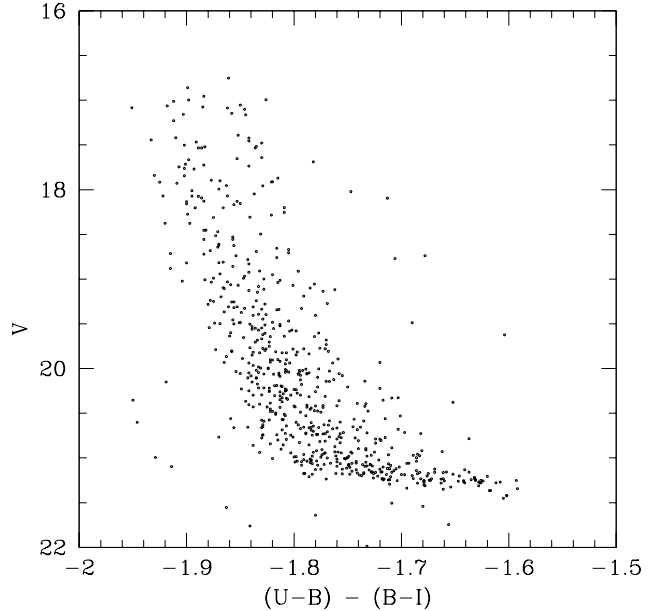}
   \caption{ The $(V, c_{UBI})$ pseudo color-magnitude diagram.  The stars in this plot were selected to lie close to the cluster RGB, as shown in Figure 7, thus suppressing field stars and cluster AGB stars.   Stars close to the cluster RGB tip are not included as they cannot photometrically be reliably separated from AGB stars.  A bifurcation of the RGB sequence is apparent.}
  \label{fig-nine}
\end{figure}

The resulting stellar sample is then plotted in the \vv vs. $c_{UBI}$ diagram (Figure 9), which as shown by \citet{mo13} rather cleanly separates multiple populations on the RGB. Their Figure 13 shows histograms of the $\Delta c_{UBI}$ parameter for 22 GC covering a wide range of metallicity, and while for the more metal-rich and metal-intermediate clusters the separation between first and second generation (1G, 2G) stars is quite apparent, the separation for the most metal-poor clusters is less obvious. This is readily explained by the general weakening, for the metal-poor stars, of the spectral signatures that drive the pseudo-color differences \citep{sb11,ca13}.   For NGC 5824 we see that the overall spread in pseudo-color for a given \vv magnitude is very similar to the  clusters of similar metallicity (e.g. NGC 6397) shown by \citet{mo13} and ascribed to a spread in light-element abundance, and the bifurcation is quite apparent.   This is best demonstrated by forming the $\Delta c_{UBI}$ parameter, following the procedure described by \citet{mo13} and as shown in Figure 10.  The main peak is populated by 1G stars, while the 2G peak is significantly less populated, the multi-Gaussian fit indicates that 75\% of the stars are 1G and 25\% are 2G. 

 It should be emphasized that all the stars plotted in Figures 9 and 10 are by selection fainter than the AGB clump (i.e. $V >17.0$) and the color-color selection strongly removes field stars, thus this sample of RGB stars is very genuine.   If we proceed to compare this result with the [Mg/Fe] distribution found by \citet{r2} for upper RGB stars we do not have such a clear-cut situation.   We have deliberately excluded such stars from Figures 9 \&10 because of the difficulty of cleanly separating RGB and AGB stars, and indeed from our photometry a few of their 26 stars fall closer to the AGB sequence than to the RGB.  In addition, 14 of the stars are nearer to the cluster center than our 90 arcsec cutoff thus contamination is a possibility, although in mitigation these are all very bright stars and photometric errors  are small.   Finally, we do not have $U$ band photometry for two stars so cannot compute the $c_{UBI}$ index for these.  We conclude that a comparison between the spread in [Mg/Fe] on the upper RGB and our photometry cannot be made definitively.   

\begin{figure}
  \includegraphics[width=\columnwidth]{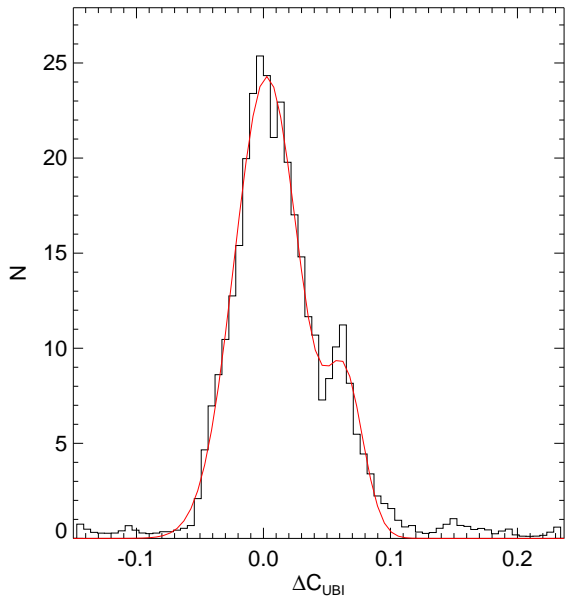}
 \caption{Histogram of the $\Delta c_{UBI}$ parameter,   This parameter is formed by first fitting a fiducial to the RGB in  the $(V, c_{UBI})$ pseudo color-magnitude diagram (Figure 9), here we linearize with $c_{UBI} = 0.00514 V^2 -0.160 V - 0.669. $  Then $\Delta c_{UBI}$ is formed for each star by measuring the difference between its pseudo-color and the fiducial.  A multi-Gaussian fit to the histogram is also shown.}

  \label{fig-ten}
\end{figure}

\section{NGC 5824 RRL Variable Stars}

The first variable star study of NGC 5824, by \citet{r1}, used the 74-inch telescope at the Radcliffe Observatory near Pretoria, South Africa
to discover 27 variables,  all of which were classified as being of RRL type with the exception that the rather brighter V11 was stated to be of uncertain classification.  From these observations only eight stars have reliable periods, and in addition the \citet{r1}  photographic photometry has an arbitrary zeropoint.  
These early results  are listed in the on-line catalog of RRL in Galactic GCs (see \citealt{cl}), where the positions for the NGC 5824 RRL were determined by \citet{sa}.\footnote{For V16 we find a position mismatch with that given by \citet{cl}, derived from \citet{sa}.  We find that our position of RA $15^h 03^m 59\overset{s}{.}31$  DEC $-33^d 05^m 08\overset{s}{.}8$ is 13.5 arcsec west and 7.8 arcsec south of the \citet{sa} position, and correctly corresponds to the star marked as V16 on Plate 1 of \cite{r1}.  We find no variable star near the alternate position, and note that positions for the other 26 variables in common agree to better than a few arcsec.}

Using the method developed by \citet{ws} and as further described by \citet{st1}, a total of 95 potential variables were flagged, of these 57 were in the vicinity of the HB and the remainder were in the BS region of the CMD.    These were examined for periodicity (see \citet{st2} for a description of the methods used) and the more difficult cases were checked multiple times using different codes and with various  selections of the input data, with cuts on photometric error and the DAOPHOT parameters $\it{chi}$ and $\it{sharp}$.    Given the long time base and many epochs of observations, for stars with stable light curves and good photometry it is possible to determine periods to close to one part in a million, so we quote periods to 6 decimal places in Table 4.   For stars with unstable light curves, or poor photometry, or poor phase cover, we have at times  had to choose one from several possible peaks in the periodogram.  Such cases are noted in the Appendix, where detailed notes on the variables can be found.  In addition, we have concentrated on finding short period variables in this study and so any variable stars  with periods longer than  a few days (e.g. RGB variables) are likely to have been missed.

\begin{table*}
\scriptsize	
 \setlength\extrarowheight{-3pt}
 \caption{NGC 5824 Variable Stars: Periods, photometric parameters, and J2000 positions}
    \label{tab:three}
    \begin{tabular}{ccccccccccccc}
    \hline
    \hline    \\
    ID & Type & Period & $<B>$ & $<V>$ & $<R>$ & $<I>$ & $A_B$ & $A_V$ & $A_R$ & $A_I$ & RA & Dec  \\
         &  &  (days) &   & & & & & & & & (hr:min:sec) & (deg:min:sec) \\
    \hline    \\
 V1  & RRab      &  0.597259 &   18.80 &   18.37 &   18.06 &   17.77 &    1.42 &    1.07 &    0.92 &    0.68 &   15 03 52.72   & -33 03 35.5 \\
V2  & RRab     &  0.649413 &   18.85 &   18.39 &   18.16 &   17.75 &    1.33 &    1.06 &    0.72 &    0.60 &   15 03 59.21   & -33 02 15.3 \\
V3  & RRab     &  0.732079 &   18.90 &   18.30 &   17.97 &   17.60 &    0.95 &    0.79 &    0.66 &    0.47 &   15 04 08.42   & -33 03 32.6 \\
V4  & RRc     &  0.334383 &   18.80 &   18.40 &   18.14 &   17.86 &    0.66 &    0.41 &    &    0.28 &   15 04 13.47   & -33 02 49.2 \\
V5  & RRab     &  0.630958 &   18.88 &   18.37 &   17.91 &   17.71 &    1.17 &    1.02 &    0.81 &    0.68 &   15 03 57.54   & -33 02 20.8 \\
V6  & RRc     &  0.307015 &   18.78 &   18.45 &   18.19 &   17.94 &    0.46 &    0.41 &    0.33 &    0.25 &   15 04 06.74   & -33 04 36.7 \\
V7  & RRc     &  0.353181 &   18.73 &   18.40 &   18.14 &   17.86 &    0.49 &    0.31 &    0.31 &    0.24 &   15 03 55.96   & -33 05 17.8 \\
V8  & RRab     &  0.601377 &   18.93 &   18.40 &   18.18 &   17.74 &    0.78 &    0.69 &    &    0.42 &   15 03 58.09   & -33 05 18.0 \\
V9  & RRab     &  0.677626 &   18.88 &   18.36 &   17.98 &   17.69 &    1.24 &    1.02 &    0.75 &    0.62 &   15 04 04.83   & -33 02 53.7 \\
V10 & RRab     &  0.694725 &   18.86 &   18.43 &   18.10 &   17.69 &    0.94 &    0.77 &    0.56 &    0.44 &   15 04 11.25   & -33 05 53.6 \\
V11 & Cep  &  2.372129 &   18.00 &   17.41 &   16.97 &   16.57 &    0.77 &    0.66 &    0.56 &    0.47 &   15 03 57.89   & -33 04 59.8 \\
V12 & RRab     &  0.580085 &   18.88 &   18.43 &   18.13 &   17.79 &    1.51 &    1.21 &    0.95 &    0.72 &   15 03 52.90   & -33 04 50.1 \\
V13 &  RRc    &  0.3675: &   18.90 &   18.40 &  18.20  &   17.86 &    &     &    &     &   15 04 00.20   & -33 05 49.5 \\
V14 & RRab     &  0.710773 &   18.88 &   18.37 &   18.01 &   17.68 &    0.94 &    0.75 &    0.65 &    0.53 &   15 04 00.18   & -33 03 17.4 \\
V15 & RRc     &  0.312907 &   18.79 &   18.46 &   18.20 &   17.96 &    0.45 &    0.37 &    0.39 &    0.25 &   15 04 05.28   & -33 04 59.8 \\
V16 & RRc     &  0.390538 &   18.78 &   18.40 &   18.08 &   17.78 &    0.47 &    0.37 &    0.32 &    0.22 &   15 03 59.31   & -33 05 08.8 \\
V17 & RRc     &  0.338638 &   18.80 &   18.43 &   18.22 &   17.87 &    0.62 &    0.45 &    0.37 &    0.28 &   15 04 01.56   & -33 05 34.8 \\
V18 & RRab     &  0.640359 &   18.87 &   18.31 &   18.15 &   17.70 &    1.18 &    0.95 &    0.85 &    0.62 &   15 04 09.17   & -33 04 07.6 \\
V19 & RRab     &  0.633701 &   18.78 &   18.22 &   18.00 &   17.61 &    1.15 &    1.01 &    0.83 &    0.58 &   15 03 56.41   & -33 04 49.0 \\
V20 & RRc     &  0.398421 &   18.78 &   18.38 &   18.07 &   17.76 &    0.40 &    0.25 &    0.35 &    0.23 &   15 03 52.34   & -33 04 30.4 \\
V21 & RRc   &  0.34353 &   18.86 &   18.42 &   18.18 &   17.83 &    &     &     &     &   15 04 02.23   & -33 02 56.4 \\
V22  & RRab   &  0.590826 &   18.89 &   18.43 &   17.95 &   17.69 &    1.11 &    0.83 &    0.70 &    0.51 &   15 04 02.53   & -33 04 20.3 \\
V23  & RRab   &  0.628086 &   18.96 &   18.43 &   18.08 &   17.76 &    1.18 &    1.04 &    0.70 &    0.67 &   15 03 49.59   & -33 08 11.9 \\
V24 & RR   &   &   18.81 &   18.39 &   18.27 &   17.86 &    &     &     &     &   15 04 07.01   & -33 09 07.5 \\
V25 & RRc     &  0.330884 &   18.82 &   18.49 &   18.23 &   17.96 &    0.63 &    0.50 &    0.43 &    0.29 &   15 03 32.38   & -33 04 08.0 \\
V26 & RRab     &  0.592856 &   18.98 &   18.45 &   18.24 &   17.74 &    1.65 &    1.37 &     &    0.64 &   15 04 29.61   & -32 57 53.6 \\
V27 & RRc     &  0.269454 &   18.73 &   18.36 &   18.14 &   17.84 &    0.59 &    0.45 &    0.36 &    0.23 &   15 04 24.58   & -33 04 20.3 \\
V28  & EB?   &              &   20.06 &   19.00 &  &   17.74 &     &    &     &     &   15 02 33.08   & -33 14 20.3 \\
V29  & RRc      &  0.304056 &   18.85 &   18.45 &   &   17.96 &    0.44 &    0.41 &     &    0.24 &   15 02 42.24   & -33 05 52.4 \\
V30 & EB       &  0.238853 &   20.17 &   19.06 &   18.21 &   17.52 &    &    &    &     &   15 03 02.09   & -33 01 05.7 \\
V31  & EB       &  0.380935 &   17.93 &   17.41 &    &   16.69 &    0.23 &    0.23 &    &    0.23 &   15 03 08.31   & -33 19 05.2 \\
V32  & RRab      &  0.614798 &   18.86 &   18.40 &   17.72 &   17.72 &    1.20 &   0.95  &   0.81 &    0.72 &   15 03 21.63   & -33 12 58.4 \\
V33  & EB?      &            &   19.78 &   18.76 &    &   17.31 &    &    &     &     &   15 03 27.58   & -33 23 07.4 \\
V34 & RRab      &  0.796201 &   18.78 &   18.33 &   17.96 &   17.57 &    0.72 &    0.62 &    0.54 &    0.35 &   15 03 42.74   & -33 00 55.0 \\
V35  & RRc      &  0.374213 &   18.81 &   18.42 &   18.08 &   17.82 &    0.44&    0.28 &    0.24 &    0.21 &   15 03 48.43   & -33 01 51.9 \\
V36 & RRc      &  0.318002 &   18.85 &   18.47 &   18.22 &   17.96 &    0.66 &    0.51 &    0.38 &    0.32 &   15 03 49.76   & -33 06 14.0 \\
V37  & RRc      &  0.410966 &   18.86 &   18.34 &   18.08 &   17.78 &    0.58 &    0.26 &    0.25 &    0.26 &   15 03 53.46   & -33 02 49.6 \\
V38  & RRc      &  0.315182 &   18.74 &   18.44 &   18.17 &   17.93 &    0.33 &    0.29 &    0.23 &    0.24 &   15 03 56.26   & -33 02 54.2 \\
V39  &  RR    &            &   18.88 &   18.27 &   18.16 &  17.57 &     &     &     &     &   15 03 58.35   & -33 04 35.9 \\
V40 & RRab      &  0.530783 &   18.91 &   18.10 &   18.15 &   17.59 &    1.54 &    1.21 &    0.96 &    0.67 &   15 03 58.90   & -33 04 33.2 \\
V41  & Cep   &  2.093630 &   17.99 &   17.35 &   16.93 &   16.48 &    0.57 &    0.43 &    0.34 &    0.27 &   15 04 00.23   & -33 04 35.2 \\
V42  & RRc      &  0.329534 &   18.78 &   18.47 &   18.16 &   17.93 &    0.59 &    0.51 &    0.41 &    0.25 &   15 04 01.98   & -33 06 56.1 \\
V43  & RRc      &  0.405010 &   18.89 &   18.39 &   18.40 &   17.79 &    0.34 &    0.15 &   &    0.16 &   15 04 03.55   & -33 03 46.1 \\
V44 & RRab      &  0.664606 &   19.03 &   18.51 &   18.14 &   17.76 &    0.63 &    0.41 &    0.39 &    0.26 &   15 04 03.90   & -33 06 57.3 \\
V45 & RRc      &  0.302787 &   18.75 &   18.43 &   18.20 &   17.94 &    0.40 &    0.36 &    0.27 &    0.22 &   15 04 16.78   & -33 02 37.5 \\
V46  & RRc      &  0.345842 &   18.77 &   18.42 &   18.23 &   17.88 &    0.56 &    0.23 &    0.30 &    0.27 &   15 04 31.72   & -33 07 02.5 \\
V47  & SXPhe &  0.037659 &   19.24 &   18.89 &    &   18.41 &    0.14 &    0.12 &     &    0.06 &   15 03 04.57   & -33 13 36.9 \\
V48  & RRc      &  0.393885 &   20.15 &   19.86 &    &   19.45 &    0.38 &    0.29 &    &    0.26 &   15 03 51.58   & -33 17 48.2 \\
V49 &    RRab      &  0.637410 &   18.85 &   18.26 &   17.95 &   17.48 &    0.85 &    0.41 &    &   &   15 03 55.88   & -33 03 34.7 \\
V50    & RRc      &  0.380136 &   18.87 &   18.48 &   18.21 &   17.89 &    0.54 &    0.45 &    0.38 &    0.31 &   15 03 55.98   & -33 04 09.8 \\
V51    & RRab      &  0.599530 &   18.76 &   18.25 &   17.86 &   17.61 &    1.27 &    1.21 &    1.16 &    0.68 &   15 03 56.82   & -33 04 01.9 \\
V52   & RRc?     &   &  18.87 &   18.38 &   18.24 &   17.76 &    &     &    &     &   15 03 58.67   & -33 03 36.8 \\
V53   & RRc      &  0.315564 &   18.55 &   18.19 &   17.99 &   17.66 &    0.51 &    0.30 &   &   0.12 &   15 04 00.62   & -33 03 57.5 \\
V54   &  EB?     &             &   18.52 &   18.09 &   18.14 &   17.49 &     &     &     &     &   15 04 38.20   & -32 50 26.0 \\
V55     & SXPhe&  0.142629 &   18.97 &   18.14 &     &   17.07 &    0.14 &    0.13 &     &    0.16 &   15 04 45.77   & -33 12 21.2 \\
V56   &   RRc      &  0.304147 &   18.69 &   18.33 &   18.13 &   17.78 &    0.41 &    0.37 &    0.29 &    0.20 &   15 03 56.45   & -33 03 57.5 \\
V57    &  RRc      &  0.314116 &   18.79 &   18.39 &   18.17 &   17.84 &    0.57 &    0.48 &    0.40 &    0.32 &   15 03 58.08   & -33 03 37.5 \\
V58  &  SXPhe        &  0.056937 &   21.12 &   20.69 &   20.41 &   20.13 &    0.31 &    0.29 &    0.18 &    0.13 &   15 03 49.59   & -33 01 17.9 \\
V59  & SXPhe        &  0.108174 &   20.42 &   19.96 &   19.65 &   19.33 &    0.30 &    0.27 &    0.15 &    0.15 &   15 04 00.28   & -33 05 35.8\\
 \hline
    \end{tabular}
    \end{table*}

The intensity-averaged magnitudes and amplitudes were obtained by fitting the light curves with templates partly based on the set of \citet{Layden1999} following the method described in \citet{ber09}.  The mean photometry is presented in Table 4 where stars V1-27 are those discovered and named by \citet{r1}, while the remainder are new discoveries here.   An example of typical phased light curves for four variables is shown in Figure 11. The high cluster star density significantly affects the photometry for distances less than an arcmin from the cluster center, and there are likely to be unidentified RRL inside that distance.
  
We have identified 47 RRL and determined periods for 44 of them, although for two stars we consider the periods to be uncertain due to poor quality photometry.

  All 26 RRL in the \citet{cl} catalog were identified and other types of variables that were discovered are discussed below.    The remaining stars in our original sample were not confirmed as periodic variables.

\begin{figure*}
  \includegraphics [scale=0.7] {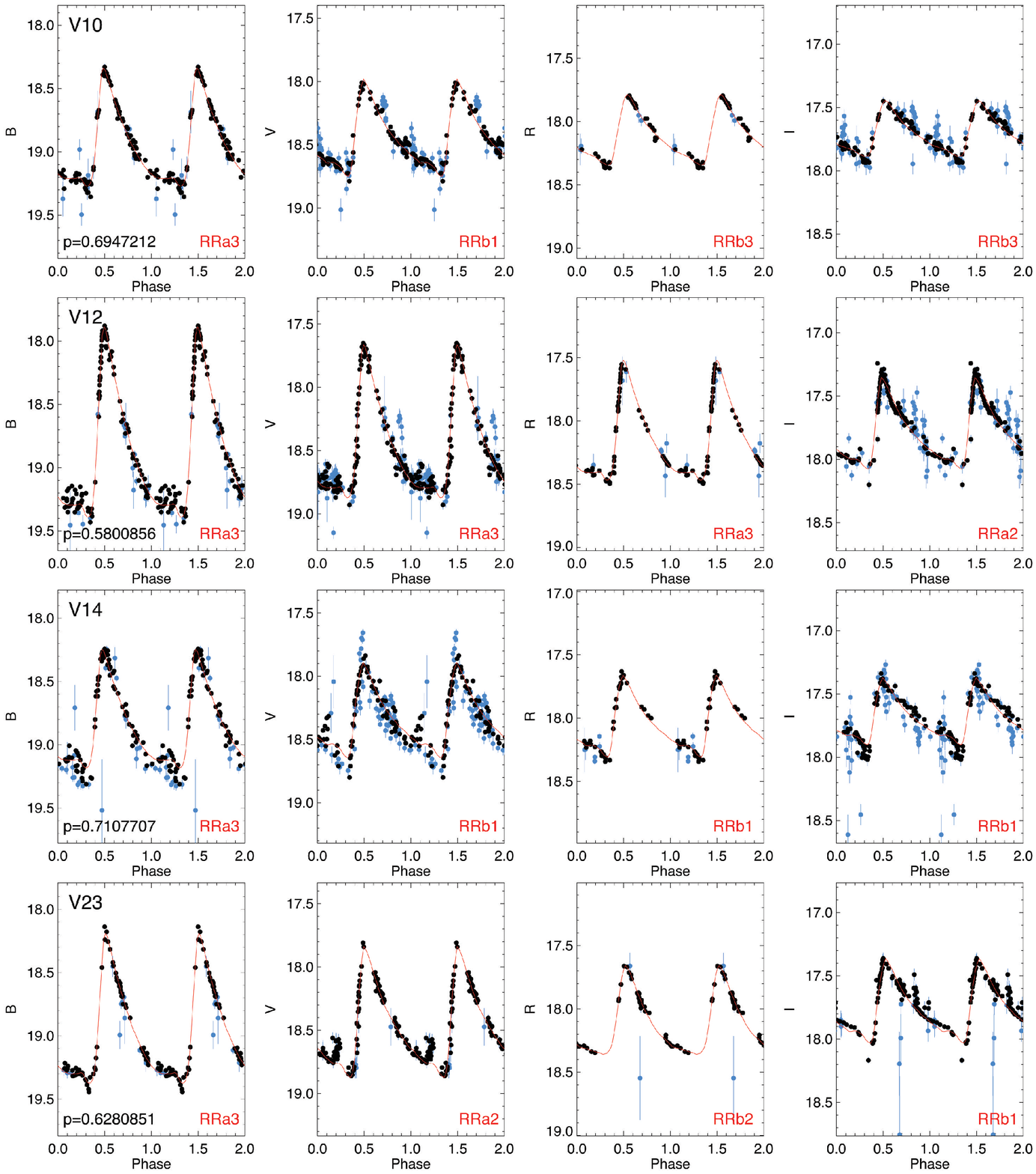}
  \caption{Phased $BVRI$ (from left to right) light curves for the RRL stars: 
V10, V12, V14 and V23 (from top to bottom). The period (in days) is given in the 
lower-left hand corner of the left panels, the name of the star in the 
upper-left hand corner of the left panels and the template fits used 
(red curves) are shown in the lower-right hand corner of each panel. 
Black circles are the data points used to obtain the template fits, 
while blue circles show the bad data points, i.e., with errors larger 
than 3$\sigma$ above the mean error of a given phased light curve; 
these were not used in the calculation of the pulsation properties.}
 \label{fig-eleven}
 \end{figure*}

\subsection{Oosterhoff Classification}

\citet{O39} showed that the mean periods of the RRL, and the ratio between the number of RRc and RRab variables divided clusters naturally into two groups, denoted OoI and OoII.  This dichotomy, originally shown for only five clusters, is true in general for Galactic GC but not for clusters in nearby galaxies, see \citet{cat09, brag16} for recent discussions;  so it is interesting to see where NGC 5824 lies in this respect.    

For the 42 NGC 5824 RRL with reliable periods, 18 of these are RRab and 24 RRc, with mean periods of $0.640 \pm 0.013$ (s.e) and $0.339 \pm 0.008$ (s.e.)  days respectively.    The ratio of RRc to RRab stars for a metal poor cluster such as NGC 5824 is expected to be higher than we find, however this is almost certainly a selection effect in that the RRab stars with high amplitudes are more likely to be discovered in the crowded central regions of the cluster than are RRc stars.  This can be demonstrated by considering the number of variables found inside a radius of 1 arcmin, where we find equal numbers of each type (6 each of RRab and RRc), so proportionally less RRc than at larger radii.  We find only two variables inside a radius of 30 arcsec, for which crowding is severe.  

Figure 12 plots the stars in the Bailey diagram, The RRL naturally group in this diagram into the fundamental mode RRab stars and the first-overtone RRc stars.    Four RRab stars (V8, V22, V40, V49) are marked as red circles and their light curves show that these stars appear to show the Blazhko effect \citep{bla07} that can cause the mean amplitude of the template fitted to all the data to be reduced in amplitude (see e.g. \citealt{ccc}).   Star V8 has a very noisy light curve and its $V$ amplitude may be as much as 0.2 mag underestimated, by comparison with the B band light curve.  Stars V22 and V49 may also have amplitudes higher than shown, however V44 has a very mild Blazhko effect, seen clearly only in the $B$ band, with the $V$ light curve relatively clean.   With the exception of these stars the remainder of the RRab stars closely follow the OoII line and the scatter about the line, and the mean period, is typical.   Star V40 has a period of 0.53 days that is 0.05 days shorter than the next shortest period RRab star.   This star has a messy light curve and it was difficult to find the period, however it does not display the Blazhko effect.  With a $V$ amplitude of 1.21 mag it lies midway between the OoI and OoII fiducials in Figure 12.   In summary the RRab stars, apart from those subject to the Blazhko effect, appear to follow the OoII fiducial and there is no evidence here for a substantial OoI component amongst the NGC 5824  RRab variables.    Many of the RRc stars in Figure 12 have poor light curves and thus are marked with red circles.  Some of these stars may be double-mode (RRd) variables, however we have not been able to determine secondary periods for any of them.   The stars with good light curves (blue circles) match the fiducial shown for M22 quite well, with a displacement of $\sim 0.02$ in log P.   This may correspond to the small difference in composition between M22 and NGC 5824.
 
\begin{figure}
   \includegraphics[width=\columnwidth]{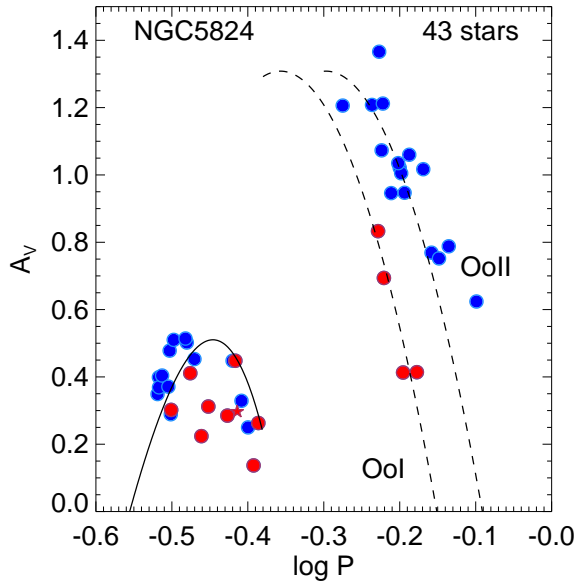}
   \caption{Bailey (or period-amplitude) diagram for the sample of RRL stars
found in NGC 5824 with reliable periods (43 stars). Red circles are those stars with 
poor-quality light curves; some or all of these stars may be displaying the Blazhko effect (if RRab) or else be double mode pulsators (if RRc). The star V48, much fainter than the HB of NGC 5824, 
is shown by a red star symbol. The dashed lines are the relations for RRab stars
in OoI and OoII clusters obtained by \citet{ccc}.
The solid curve is derived from the M22 (OoII cluster) RRc variable study by \citet{ku13b}.}
 \label{fig-twelve}
\end{figure}

\subsection {Distance}

We have calculated the distance to NGC 5824 using the period-Wesenheit relations for the indices (V, B-V) and (V, B-I). Both relations can be considered metallicity-independent \citep{mar15, ma15}.    We applied here one of the methods (so-called "theoretical", \citealt{ma15}) to calculate the distance, using the two given relations applied to three different samples: RRab, RRc and RRab+RRc, where the RRc periods have been fundamentalized by adding 0.127 to log Period \citep{ib74, mar03}.   The estimated distance modulus is $\mu_0 = 17.45 \pm 0.07 $(systematic) $\pm 0.02 $(random), where the mean random error per star  is 0.13 mag.  This corresponds to a distance of 30.9 Kpc.   The fits are shown in Figure 13.   There is some tension between this result and that from the isochrone matching (17.58) but formally they are just in agreement, given the errors.

\begin{figure*}
   \includegraphics[scale=0.8]{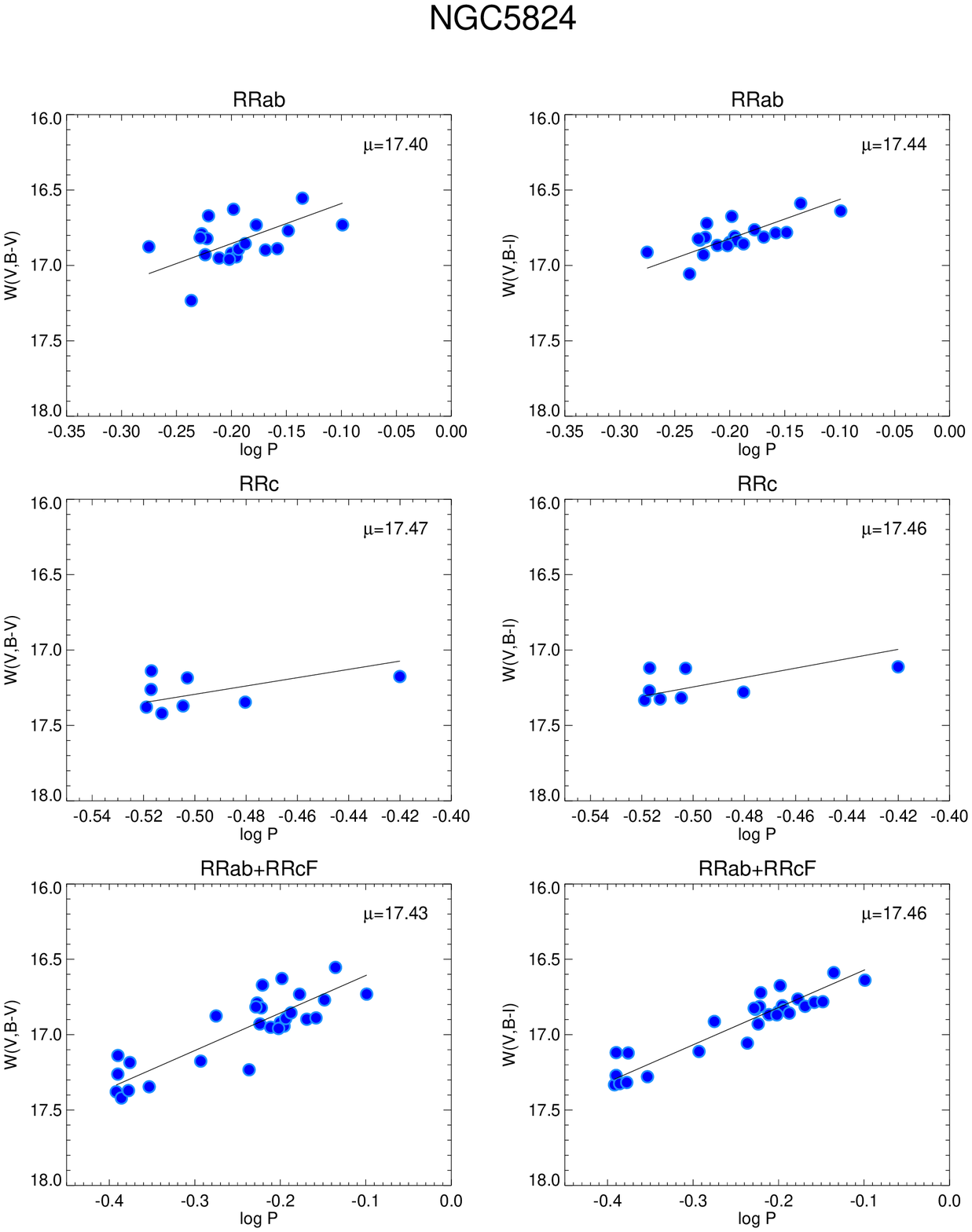}
   \caption{ Apparent Wesenheit magnitudes -- $V$,$B$-$V$ (left) and the $V$, $B$-$I$ 
(right) -- versus the logarithm of the period for the RRL stars of NGC 5824. The black lines
are the theoretical period-Wesenheit relations \citep{mar15,ma15} shifted an amount equal to the distance modulus (labeled in the upper-right hand corner
of each panel) derived from each subsample grouped as (from top to bottom) RRab, RRc, RRab 
plus RRc with periods fundamentalized (RRcF). See the text for further details.}
   \label{Fig-thirteen}
\end{figure*}

\section{Other variable stars}

Two variable stars with periods just over two days  were found lying $\sim 1$ mag above the red side of the HB.  Stars in this general position of the CMD, assuming they are cluster members, can be either Type II Cepheids or Anomalous Cepheids (AC).  Short period ($1-5$ days) Type II Cepheids are called BL Herculis stars and are rapidly evolving from the HB at the end of core-helium burning towards the AGB.    The typical lifetime of a star in this phase of evolution is $\sim 8\e{5}$ years, about $1\%$ of the lifetime of the HB phase \citep{cs13}.  Thus, it is expected to find only a few such stars in a given GC, even those with a richly populated HB.  The AC are metal-poor stars  of $1-2$ solar masses, and in the context of NGC 5824 they are the progeny of the blue stragglers, themselves mergers of primordial close binaries and/or products of collisions in the dense GC core.   Only one AC is known in a GC \citep{n94}. 

 \citet{so15}  show (their Fig 3) that the light curve shapes can be useful in discriminating between the various classes of short-period pulsating variables, and in particular the $I$ band light curves for BL Her stars demonstrate a characteristic rapid rise to almost maximum light followed by a rounded rather than sharp peak at maximum light.  This is very similar to the light curve shape for V11 in the redder passbands.  AC light curves have a more classical saw-tooth appearance and V41, with noisier data and some phase gaps, cannot be so easily classified as V11 but may be an AC from this criterion alone.  The extensive discussion of population II variables by \citet{n94} suggests that stars with periods longer than 2 days and luminosities $\sim 3$ times that for the RRL are probably Type II Cepheids and not ACs.   \citet{so15} confirm this in their Figures 5 and 6 which show the $I$-band period-luminosity (PL) relation  for short period variables in the LMC and SMC.  Their plots show that only 7 out of 250 AC have periods over 2 days, and that the luminosity separation between the two classes of variable appears to be a weak function of metallicity,  important since the NGC 5824 variables are one dex more metal-poor than the SMC average.  V11 lies very close to the Type II Cepheid PL relation, however V41, slightly brighter and wth shorter period,  lies close to midway between the two PL relations and so this test cannot be considered definitive for V41, however the rather close position of the two stars in the CMD and the presence of several supra-HB stars (see section 3.3) would lead to the conclusion  that both V11 and V41 are Type II Cepheids, but that  the possibility of V41 being an AC cannot be completely ruled out.
 
\begin{figure}
   \includegraphics[width=\columnwidth]{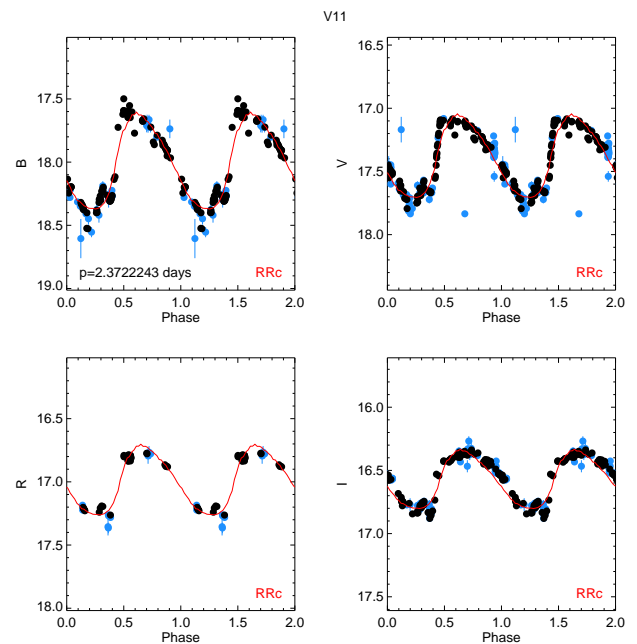}
   \caption{Phased $BVRI$ light curves for the variable star V11, period 2.373 days, which is a 
probable Type II Cepheid. See caption of Figure 11 for details.}
 \label{fig-fourteen}
\end{figure}

\begin{figure}
   \includegraphics[width=\columnwidth]{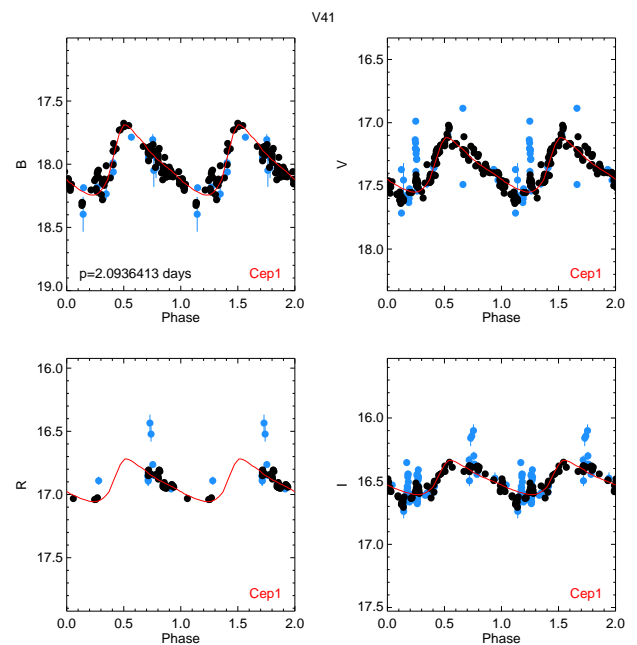}
   \caption{Phased $BVRI$ light curves for the variable V41, period 2.093 days, which is probably a Type II Cepheid although 
classification as an Anomalous Cepheid is also possible.   See caption of Figure 11 for details. }
  \label{fig-fifteen}
\end{figure}
 
V48 (W46088) is a magnitude fainter than the cluster RRL and near the blue edge of the instability strip,  it has an RRc-like light curve and a tentative period of 0.393885d, the phased light curve is rather noisy and this period may not be the correct one.  NGC 5824 variable stars in this region of the CMD would be SX Phe stars with much shorter periods.  V48 is quite distant from the cluster center at 13.7 arcmin and we interpret this star as a background Galactic halo RRL with a distance modulus of $19.42 \pm0.12$, $\sim77$ Kpc  using the same methods as described above for the cluster RRL.  The most extensive sample of distant Galactic RRL candidates is found in \citet{he16}, with stars covering a range in distance from $10-120$ kpc, so while W46099 is more distant than most known Galactic halo RRL it is not the most distant.  In any case, as pointed out by these authors and others, there are likely many more distant Galactic RRL waiting to be discovered.

SX Phe stars in globular clusters are pulsating blue stragglers, these stars have periods much shorter than the RRL.  We identify four such stars here: V47, V55, V58, V59.   SX Phe stars follow a steep PL relation (see discussions in \citealt{mc11, fi15}), with parallel sequences for fundamental and first overtone pulsators, and from their positions in the PL diagram it seems likely that V47 is a first overtone pulsator and the remaining stars are pulsating in the fundamental mode. 

Finally, we have determined periods for two eclipsing binaries (V30, V31) and three other stars (V28, V33, V54) for which we have not been able to determine periods are also likely to be eclipsing binaries.  All these stars are more distance than 17 arcmin from the center of NGC 5824 and also fall well off the cluster sequences, and thus are classified as field stars.

\section{Discussion and Conclusions}

\citet{mil16} compellingly show from high precision RGB photometry for 57 GC in bands including the vacuum-UV that 1G and 2G stars clearly separate in pseudo two-color diagrams.  Unfortunately NGC 5824 is not one of the  clusters included in this study.  \citet{mil16} further show that although most clusters exhibit only 1G and 2G  sequences, cleanly separated in their diagrams (denoted Type I), some clusters, particularly the more massive ones, show merged or multiple 1G and/or 2G sequences (denoted Type II).  Most if not all Type II clusters also exhibit multi-sequence behavior in optical CMDs indicative of a more complex star formation history than the Type I clusters, for instance NGC 1851 (\citealt{cu17} and references therein) has multiple SGB indicative of C+N+O and/or He variations and complex MS, RGB and HB sequences.   Given that NGC 5824 is a rather massive cluster (see Introduction) it is of interest to see where it fits in this scenario.   The $c_{UBI}$ diagram (Figures 9 \& 10) shows that NGC 5824 has a spread in pseudo-color, with double peak, similar to other GC of similar low metallicity \citep{mo13} and indicative of  a mixture of 1G (75\%) and 2G (25\%) stars.  There is no sign of other stellar populations and so this would indicate that NGC 5824 is a Type I cluster, however precision photometry of the MS and SGB stars with much higher spatial resolution than here - with HST or adaptive optics on a large ground-based telescope - would permit a stronger statement to be made.   Additionally,  
a high-dispersion abundance study that includes the light elements (e.g. O, Na, Al) would more clearly decipher the chemical enrichment history of the 1G and 2G stars, and indicate whether there are other star-star abundance differences that would indicate a more complex evolutionary history than the simple picture outlined here.

While the presence of RRL is well known to indicate that an old population $10-13$ Gyr is present, (e.g. \citealt{w89} and references therein) 
the pulsational properties of the variables together with the morphology of the HB can further illuminate the star formation history and element enrichment during that period, see e.g \citet{ku13,mi14,ma15,fi16}.  We show above via the Bailey diagram (Figure 12) in particular that the RRL define a pure OoII population with little spread in metallicity, a result that is consistent with the narrow RGB and the appearance of the HB.    This evidence would further support the classification of NGC 5824 as a Type I cluster, with star formation quenched very early.

Turning now to NGC 5824's relationship to its environment, as discussed above this study confirms earlier results but is sufficiently spatially inhomogeneous so as not to be able to make an advance on work already in the literature.   While there appears to be no direct association of NGC 5824 with known stellar streams,  wider and deeper surveys are showing that tidal disruption tails associated with globular clusters may be ubiquitous \citep{ber16}.   Although the high Galactic foreground star density makes such a detection difficult for NGC 5824, a study of the star density around the cluster on a scale of a few square degrees may be rewarding.

We conclude  that NGC 5824 appears to be a relatively normal Oosterhoff Type II globular cluster and there is no evidence from this data set indicating that it had an exotic past, such as now being the remnant nucleus of a dwarf galaxy that has been completely disrupted by tidal interaction with our Galaxy.

\section*{Acknowledgments}

CEMV and MM acknowledge support from the IAC (grant P/301204) and from the Spanish Ministry of Economy and Competitiveness (grant AYA2014-56795).  SC warmly acknowledges partial financial support from PRIN-INAF2014 (PI: S. Cassisi).  This research draws upon data obtained through the facilities of the Canadian Astronomy Data Centre operated by the National Research Council of Canada with the support of the Canadian Space Agency; data obtained from the ESO Science Archive Facility under multiple requests by the authors; data obtained from the Isaac Newton Group Archive, which is maintained as part of the CASU Astronomical Data Centre at the Institute of Astronomy, Cambridge; and data distributed by the Science Data Archive at NOAO. NOAO is operated by the Association of Universities for Research in Astronomy (AURA) under a cooperative agreement with the National Science Foundation.  This project also  used data obtained with the Dark Energy Camera (DECam), which was constructed by the Dark Energy Survey (DES) collaboration.  Funding for the DES Projects has been provided by 
the U.S. Department of Energy, 
the U.S. National Science Foundation, 
the Ministry of Science and Education of Spain, 
the Science and Technology Facilities Council of the United Kingdom, 
the Higher Education Funding Council for England, 
the National Center for Supercomputing Applications at the University of Illinois at Urbana-Champaign, 
the Kavli Institute of Cosmological Physics at the University of Chicago, 
the Center for Cosmology and Astro-Particle Physics at the Ohio State University, 
the Mitchell Institute for Fundamental Physics and Astronomy at Texas A\&M University, 
Financiadora de Estudos e Projetos, Funda{\c c}{\~a}o Carlos Chagas Filho de Amparo {\`a} Pesquisa do Estado do Rio de Janeiro, 
Conselho Nacional de Desenvolvimento Cient{\'i}fico e Tecnol{\'o}gico and the Minist{\'e}rio da Ci{\^e}ncia, Tecnologia e Inovac{\~a}o, 
the Deutsche Forschungsgemeinschaft, 
and the Collaborating Institutions in the Dark Energy Survey. 
The Collaborating Institutions are 
Argonne National Laboratory, 
the University of California at Santa Cruz, 
the University of Cambridge, 
Centro de Investigaciones En{\'e}rgeticas, Medioambientales y Tecnol{\'o}gicas-Madrid, 
the University of Chicago, 
University College London, 
the DES-Brazil Consortium, 
the University of Edinburgh, 
the Eidgen{\"o}ssische Technische Hoch\-schule (ETH) Z{\"u}rich, 
Fermi National Accelerator Laboratory, 
the University of Illinois at Urbana-Champaign, 
the Institut de Ci{\`e}ncies de l'Espai (IEEC/CSIC), 
the Institut de F{\'i}sica d'Altes Energies, 
Lawrence Berkeley National Laboratory, 
the Ludwig-Maximilians Universit{\"a}t M{\"u}nchen and the associated Excellence Cluster Universe, 
the University of Michigan, 
{the} National Optical Astronomy Observatory, 
the University of Nottingham, 
the Ohio State University, 
the University of Pennsylvania, 
the University of Portsmouth, 
SLAC National Accelerator Laboratory, 
Stanford University, 
the University of Sussex, 
and Texas A\&M University.

\software{DAOPHOT/ALLSTAR/ALLFRAME \citep{st87, st94}}

\section{Appendix 1 - Notes on individual variable stars}  

Earlier work can be found in \citet{r1} and \citet{cl}, who provides positions determined by \citet{sa}.  For the new variable star discoveries  (V28-V59) we include the internal identification in our catalog.\\
V1:  Rosino's period of 0.597d is confirmed.\\
V2:  Rosino's period of 0.651d is close to that found here.  The $R$ band maximum is missing.\\
V3:  Rosino's period of 0.641d is not confirmed, we find 0.732d.  Possible Blazhko star.\\
V4:  The period is well-determined, but there is systematic scatter between the sets of observations.  Possibly RRd, however a definite secondary period could not be found.   The scatter remained even after very tightly selecting the lowest error observations, and the star is not crowded.\\
V5:  Rosino's period of 0.634d is confirmed. $R$ band maximum is missing.\\
V6:  Well defined light curve in all colors.\\
V7:  Probable RRd given the light curve scatter, but the secondary period was not found reliably. \\
V8:  Noisy data  Variable period? Probable Blazhko.\\
V9:  Significant period change.\\
V10: Well-defined light curve.\\
V11:  Period is 2.372d, classified as a Type II Cepheid.  The star is 1 arcmin from the cluster center. \\
V12   Excellent light curves in all colors.   Period change visible.\\
V13:  This star seems to have period of about 0.3675d, changing rapidly as a function of time.   Alternative much shorter periods seem less likely. \\
V14:  Our period of 0.71077d is close to double the 0.35d tentatively found by Rosino. \\
V15:  Noisy light curve. \\
V16:  Considerable scatter.  \\
V17:  Possible RRd?\\
V18:  Good light curves, \\
V19:  Blazhko? \\
V20: Many outliers particularly in $B$ and $V$. There are alternate period possibilities.\\
V21: Period probably 0.343353d, several outliers.\\ 
V22: Blazhko. \\
V23:  Good light curves.\\
V24: Period possibly close to 0.33 days.\\
V25: Good light curves generally, more scatter in $I$. \\
V26: R band has poor phase cover.\\
V27: Noisy, period may be incorrect. \\
V28 = C7884:  Probable eclipsing binary, not possible to determine the period.\\
V29 = C10632:  No $R$ band data.\\ 
V30 = C16441:  Eclipsing Binary. \\
V31 = C18400: Eclipsing Binary. \\
V32 = C22641:  Good light curves. \\ 
V33 = C24689:  Probable Eclipsing Binary, not possible to determine the period.\\
V34 = C31073:  Good light curves. The longest period RRab star.\\
V35 = C34409:  Very noisy light curve, pulsational parameters uncertain \\  
V36 = C35369:  Some outliers.\\
V37  = C38462:  Some outliers.\\
V38 = C41028:    Noisy with some outliers.\\
V39 = C43242:   Probable RRc or RRd, period not determined.\\
V40 = C43909:  Need to select only the best data in order to find the period, definitely an RRab light curve,  period of 0.530783d. is 0.05d shorter than the next shortest period RRab star. \\
V41 = C45484:  Period 2.09363d, classed as Type II Cepheid.    $R$ band maximum is missing.\\
V42 = C47656:  Incomplete phase cover in $R$.\\
V43 = C49572:  Much scatter.\\
V44 = C49983:  Good light curves. Mild Blazhko effect seen clearly in the $B$ band.\\
V45 = C59386:   Good light curves.\\
V46 = C65577:  Rather noisy.\\
V47 = W19109:  SX Phe, no $R$ band data.\\
V48 = W46088:  Fainter than the cluster RRL, no $R$ band data.\\
V49 = X30:  Noisy, poor phase cover near maximum in all colors.  Probable Blazhko.\\
V50 = X31:  Light curve shows double sequence, probably due to blending in poor seeing.\\
V51 = X34:  R band lacks maximum.\\
V52 = X38:   Looks like an RRc, but period could not be definitively found.\\
V53 = X41:   RRc with very noisy light curves, only $B$ and $R$ band are reliable.  This is the closest RRL to the center of NGC 5824 (0.38 arc min) that we have identified. \\
V54 = X49:   Probable eclipsing binary.\\
V55 = X51:  SX Phe.\\
V56 = Y2:  Quite noisy.\\
V57 = Y24:  Noisy.\\
V58 = SX7:  SX Phe.\\ 
V59 = SX23:  SX Phe.\\

\label{lastpage}

\begin{thebibliography}{99}

\bibitem[\protect\citeauthoryear{Bellini et al.}{2017}]{be17}Bellini, A., Anderson, J., van der Marel, R.P., et al., 2017, arXiv:1705.01951
\bibitem[\protect\citeauthoryear{Bernard et al.}{2009}]{ber09}Bernard, E.J., Monelli, M., Gallart, C.,  et al., 2009, ApJ, 699, 1742
\bibitem[\protect\citeauthoryear{Bernard et al.}{2016}]{ber16}Bernard, E.J., Ferguson, A.M.N., Schlafly, E.F., et al., 2016, MNRAS, 463, 1759
\bibitem[\protect\citeauthoryear{Bla\u{z}ko}{1907}]{bla07} Bla\u{z}ko, S., 1907, Astron. Nachr., 175, 325
\bibitem[\protect\citeauthoryear{Bono et al.}{2013}]{bo10} Bono, G., Stetson, P.B., Walker, A.R., et al.,  2010, PASP, 122, 651 
\bibitem[\protect\citeauthoryear{Braga et al.}{2016}]{brag16} Braga, V.F., Stetson, P.B., Bono, G., et al.,  2016, AJ, 152, 170 
\bibitem[\protect\citeauthoryear{Brocato et al.}{1995}]{b1} Brocato, E., Buonanno, R., Malakhova, Y., Piersimoni, A.M., 1996, A\&A, 311, 778
\bibitem[\protect\citeauthoryear{Cacciari et al.}{2005}]{ccc}Cacciari, C., Corwin, T.M., Carney, B.W., 2005, AJ, 129, 267
\bibitem[\protect\citeauthoryear{Cannon et al.}{1990}]{c1} Cannon, R.D., Sagar, R., Hawkins, M.R.S., 1990, MNRAS, 243, 151
\bibitem[\protect\citeauthoryear{Carballo-Bello et al.}{2012}]{cb12} Carballo-Bello, J.A., Gieles, M., Sollima, A., at al., 2012, MNRAS, 419, 14
\bibitem[\protect\citeauthoryear{Carballo-Bello et al.}{2014}]{cb14} Carballo-Bello, J.A., Sollima A., Mart\'{i}nez-Delgado D., et al., 2014, MNRAS, 445, 2971
\bibitem[\protect\citeauthoryear{Carretta et al.}{2009}]{ca09} Carretta, E., Bragaglia, A., Gratton, R., D'Orazi, V., Lucatello, S., 2009, A \&A, 508, 695
\bibitem[\protect\citeauthoryear{Cassisi et al.}{2007}]{ca07} Cassisi, S., Potekhin, A.Y., Pietrinferni, A., Catelan, M., Salaris, M., 2007, ApJ, 661, 1094
\bibitem[\protect\citeauthoryear{Cassisi \& Salarisl}{2013}]{cs13}Cassisi, S., Salaris, M., 2013, In "Old Stellar Populations"  publ. Wiley-VCH, p.197
\bibitem[\protect\citeauthoryear{Cassisi et al.}{2013}]{ca13}Cassisi, S.,  Mucciarelli, A., Pietrinferni, A., Salaris, M., Ferguson, J., 2013, A\&A 554, A19
\bibitem[\protect\citeauthoryear{Catelan}{2009}]{cat09}Catelan, M., 2009, AP\&SS, 320, 216
\bibitem[\protect\citeauthoryear{Cheselka et al.}{1995}]{ch93} Cheselka, M., Holberg, J.B., Watkins, R., Collins, J., 1993, AJ, 106, 2365
\bibitem[\protect\citeauthoryear{Clement et al.}{2001}]{cl} Clement, C., Muzzin, A., Dufton, Q., et al.,  2001, AJ, 122, 2587.  With updates, access the catalog at http://www.astro.utoronto.ca/~cclement/
\bibitem[\protect\citeauthoryear{Cummings et al.}{2017}]{cu17} Cummings, J.D., Geisler, D., Villanova, S., 2017, AJ, 153, 192   
\bibitem[\protect\citeauthoryear{Da Costa et al.}{2014}]{d1}   Da Costa, G. S., Held, E.V., Savianne, I., 2014, MNRAS, 438, 3507
\bibitem[\protect\citeauthoryear{Ferraro et al.}{1991}]{f91}  Ferraro, F.R., Clementini, G., Fusi Pecci, F., Buonanno, R., 1991, MNRAS, 252, 357 
\bibitem[\protect\citeauthoryear{Fiorentino et al.}{2015}]{fi15} Fiorentino, G., Marconi, M., Bono, G., et al., 2016, ApJ, 810, 15
\bibitem[\protect\citeauthoryear{Fiorentino et al.}{2016}]{fi16} Fiorentino, G., Bono, G., Monelli, M., et al., 2016, ApJ, 798, 12
\bibitem[\protect\citeauthoryear{Guldenschub et al.}{2005}]{gu05}  Guldenschub, K.A., Layden, A.C., Wan, Y., et al., 2005, PASP, 117, 721
\bibitem[\protect\citeauthoryear{Gratton et al.}{2012}]{g1} Gratton, R.C., Carretta, E., Bragaglia, A., 2012, ARAA, 20, 50
\bibitem[\protect\citeauthoryear{Grillmair et al.}{1995}]{g2} Grillmair, C.J., Freeman, K.C., Irwin, M., Quinn, P.J., 1995, AJ, 109, 2553
\bibitem[\protect\citeauthoryear{Harris}{1996}]{h1}  Harris, W.E., 1996, AJ, 112, 1487
\bibitem[\protect\citeauthoryear{Harris}{1975}]{h2}  Harris, W.E., 1975, ApJS, 29, 397
\bibitem[\protect\citeauthoryear{Hernitschek et al.}{2016}]{he16} Hernitschek, N., Schlafly, E.; Sesar, B., et al., 2016, ApJ, 817, 73
\bibitem[\protect\citeauthoryear{Iben}{1974}]{ib74} Iben, I., Jr., 1974, ARA\&A, 12. 215
\bibitem[\protect\citeauthoryear{Innes}{1925}]{i1} Innes R.T. A., 1925, Circular of the Union Observatory Johannesburg, 66, 328
\bibitem[\protect\citeauthoryear{Kinman}{1959}]{k1} Kinman, T.D., 1959, MNRAS, 119, 538
\bibitem[\protect\citeauthoryear{Kunder et al.}{2013a}]{ku13}Kunder, A., Stetson, P.B., Catelan, M., Walker, A.R., Amigo, P., 2013, AJ, 145, 25
\bibitem[\protect\citeauthoryear{Kunder et al.}{2013b}]{ku13b}Kunder, A., Stetson, P.B., Cassisi, S., et al., 2013, AJ, 146, 119
\bibitem[\protect\citeauthoryear{Landolt}{1992}]{lan92} Landolt, A.U., 1992,  AJ, 104, 340  
\bibitem[\protect\citeauthoryear{Law \& Majewski.}{2010}]{lm} Law, D.R., Majewski, S.R., 2010, ApJ, 718, 1128
\bibitem[\protect\citeauthoryear{Layden et al.}{1999}]{Layden1999} Layden, A.C., Ritter, L.A., Welch, D.L., Webb, T.M.A., 1999, AJ, 117, 1313
\bibitem[\protect\citeauthoryear{Lee}{1991}]{le91} Lee, Y-W, 1991, ASPC, 13, 2051 
\bibitem[\protect\citeauthoryear{Marconi et al.}{2003}]{mar03} Marconi, M.,Caputo, F., Di Criscienzo, M., Castellani, M., 2003, ApJ., 596, 299
\bibitem[\protect\citeauthoryear{Marconi et al.}{2015}]{mar15} Marconi, M., Coppola, G., Bono, G., et al., 2015, ApJ., 808, 50
\bibitem[\protect\citeauthoryear{Mart\'{i}nez-V\'{a}zquez et al.}{2015}]{ma15} Mart\'{i}nez-V\'{a}zquez, C.E., Monelli, M., Bono, G., et al., 2015, MNRAS, 454, 1509
\bibitem[\protect\citeauthoryear{McNamara}{2011}]{mc11} McNamara, D.H., 2011, AJ, 142, 100
\bibitem[\protect\citeauthoryear{Milone et al.}{2013}]{m1} Milone, A.P., Marino, A.F., Piotto, G., et al., 2013, ApJ, 767, 120
\bibitem[\protect\citeauthoryear{Milone et al.}{2014}]{mi14} Milone, A.P., Marino, A.E., Dotter, et al., 2014, ApJ, 785, 21
\bibitem[\protect\citeauthoryear{Milone et al.}{2017}]{mil16} Milone, A.P., Piotto, G., Renzini, A., et al., 2017, MNRAS, 464, 3636
\bibitem[\protect\citeauthoryear{Miocchi et al.}{2013}]{mi13} Miocchi, P., Lanzoni, B., Ferraro, F. R., et al., 2013, ApJ, 774, 151
\bibitem[\protect\citeauthoryear{Moehler et al.}{2007}]{mo2007} Moehler, S., Dreizler, S., Lanz, T., et al., 2007, A\&A, 475, L5
\bibitem[\protect\citeauthoryear{Momany et al.}{2004}]{mo04} Momany, Y., Bedin, L.R., Cassisi, S., et al., 2004, A\&A, 420, 605
\bibitem[\protect\citeauthoryear{Monelli et al.}{2013}]{mo13} Monelli, M., Milone, A. P., Stetson, P. B., et al., 2013, MNRAS, 431, 2126
\bibitem[\protect\citeauthoryear{Nemec et al.}{1994}]{n94}  Nemec, J.M., Nemec, A.F.L., Lutz, T.E., 1994, AJ, 108, 222
\bibitem[\protect\citeauthoryear{Newbury et al.}{2009}]{n1} Newbury, H.J., Yanny, B., Willett, B.A., 2009, ApJ, 700, 71
\bibitem[\protect\citeauthoryear{Oosterhoff}{1939}]{O39} Oosterhoff, P.T., 1939, The Observatory, 62, 104
\bibitem[\protect\citeauthoryear{Pawlowski et al.}{2012}]{p12} Pawlowski, M.S., Pflamm-Altenburg, J., Kroupa, P., 2012, MNRAS, 423, 1109
\bibitem[\protect\citeauthoryear{Pietrinferni et al.}{2004}]{p2} Pietrinferni, A., Cassisi, S., Salaris, M., Castelli, F., 2004, ApJ, 612, 168
\bibitem[\protect\citeauthoryear{Pietrinferni et al.}{2006}]{p4} Pietrinferni, A., Cassisi, S., Salaris, M., Castelli, F., 2006, ApJ, 642, 797
\bibitem[\protect\citeauthoryear{Pietrinferni et al.}{2013}]{p3} Pietrinferni, A., Cassisi, S., Salaris, M., Hidalgo, S., 2013, A\&A, 558, A46
\bibitem[\protect\citeauthoryear{Piotto et al.}{2002}]{P8} Piotto, G., King, I. R., Djorgovski, S. G.,  et al., 2002, A\&A, 391, 945
\bibitem[\protect\citeauthoryear{Piotto et al.}{2009}]{p1} Piotto, G., 2009, in IAU Symposium vol. 258, ed. E.E. Mamajek, D.R. Soderblom,\& R.F.G. Wyse, pp. 233-244
\bibitem[\protect\citeauthoryear{Potekhin}{1999}]{po99} Potekhin, A. Y. 1999, A\&A, 351, 787
\bibitem[\protect\citeauthoryear{Renzini et al.}{2015}]{re15} Renzini, A., D'Antona, F., Cassisi, S., et al., 2015, MNRAS, 454, 4197
\bibitem[\protect\citeauthoryear{Roederer et al.}{2016}]{r2} Roederer, I.U., Mateo, M., Bailey III, J.I., et al., 2016, MNRAS, 455, 2417
\bibitem[\protect\citeauthoryear{Rosino}{1961}]{r1} Rosino, L., 1961,   PASP, 73, 309
\bibitem[\protect\citeauthoryear{Samus et al.}{2009}]{sa} Samus, N.N., Kszarovets, E.V., Pastukhova, E.N., Tsvetkova, T.M., Durlevich, O.V., 2009, PASP, 121, 1378
\bibitem[\protect\citeauthoryear{Sandage \& Katem}{1983}]{sk83} Sandage, A., Katem, B., 1983, AJ, 88,1146
\bibitem[\protect\citeauthoryear{Sanna et al.}{2014}]{s2} Sanna, N., Dalessandro, E., Ferraro, F. R., et al., 2014, ApJ,78, 90
\bibitem[\protect\citeauthoryear{Sarajedini \& Layden}{1997}]{sl97} Sarajedini, A., Layden, A., 1997, AJ, 113, 264
\bibitem[\protect\citeauthoryear{Savianne et al.}{2012}]{s1}   Savianne, I., Da Costa, G. S., Held, E.V., et al., 2012, A\&A,540, A27
\bibitem[\protect\citeauthoryear{Sbordone et al.}{2011}]{sb11}  Sbordone, L., Salaris, M., Weiss, A., Cassisi, S. 2011, A\&A, 534, A9
\bibitem[\protect\citeauthoryear{Schlafly \& Finkbeiner}{2011}]{s3} Schlafly, E.F., Finkbeiner, D.P., 2011, ApJ, 737, 1035
\bibitem[\protect\citeauthoryear{Soszy\'{n}ski et al.}{2015}]{so15} Soszy\'{n}ski, I., Udalski, A., Szyma\'{n}ski, M. K., et al., 2013, AcA., 65, 233
\bibitem[\protect\citeauthoryear{Stetson}{1987}]{st87} Stetson, P.B., 1987, PASP, 99, 191
\bibitem[\protect\citeauthoryear{Stetson}{1994}]{st94} Stetson, P.B., 1994, PASP, 106, 250
 \bibitem[\protect\citeauthoryear{Stetson}{1996}]{st1} Stetson, P.B., 1996, PASP, 108, 851
\bibitem[\protect\citeauthoryear{Stetson et al.}{2014}]{st2} Stetson, P.B., Braga, V.F., Dall'Oro, M., et al., 2014, PASP, 121, 521 
\bibitem[\protect\citeauthoryear{Sturch}{1966}]{st66} Sturch, C., 1966, ApJ, 143, 774 
\bibitem[\protect\citeauthoryear{Trager et al.}{1995}]{tr95}  Trager, S.C.,  Djorgovsky, S.G., King, I., 1995, AJ, 109, 218
\bibitem[\protect\citeauthoryear{Walker}{1989}]{w89} Walker, A.R., 1989, PASP, 101, 570
\bibitem[\protect\citeauthoryear{Walker}{1990}]{wa90} Walker, A.R., 1990, AJ, 100, 1532
\bibitem[\protect\citeauthoryear{Walker}{1992}]{wal92} Walker, A.R., 1992, ApJ, 390, L81 
\bibitem[\protect\citeauthoryear{Welch \& Stetson}{1993}]{ws} Welch, D.L., \& Stetson, P.B., 1993, AJ, 105, 1813
\bibitem[\protect\citeauthoryear{Zinn \& West}{1984}]{zw84} Zinn, R., \& West, M.J., 1984, ApJS, 55, 45
\bibitem[\protect\citeauthoryear{Zinn}{1985}]{z1} Zinn, R., 1985, ApJ, 293, 424

\end{thebibliography}
\end{document}